\documentclass{article}
\date{}
\usepackage{multirow,color, url, graphics}
\usepackage{subfigure}
\usepackage{float}
\usepackage{epstopdf}
\usepackage{listings}
\usepackage{xcolor}
\usepackage{amsmath}

\usepackage{geometry}
\usepackage{multicol,graphicx,amssymb,mathrsfs,amsmath,pifont,amscd,latexsym,amsthm,color,fancyhdr,CJK,url,hyperref,longtable, pifont, subfigure,epstopdf,setspace,multirow,colortbl,tabularx,threeparttable, booktabs, verbatim, bbm,listings, balance}
\usepackage[linesnumbered,boxed,algosection]{algorithm2e}

\begin{document}
\title{Understanding Quality of Experiences on Different Mobile Browsers: Measurements, Analysis, and Implications}
\author{Yun Ma and Shuailiang Dong}

\maketitle
\begin{abstract}
The web browser is one of the major channels to access the Internet on mobile devices. Based on the smartphone usage logs from millions of real-world Android users, it is interesting to find that about 38\% users have more than one browser on their devices. However, it is unclear whether the quality of browsing experiences are different when visiting the same webpage on different browsers. In this paper, we collect 3-week consecutive traces of 337 popular webpages on three popular mobile browsers: Chrome, Firefox, and Opera. We first use a list of metrics and conduct an empirical study to measure the differences of these metrics on different browsers. Then, we explore the variety of loading time and cache performance of different browsers when visiting the same webpage, which has a great impact on the browsing experience. Furthermore, we try to find which metrics have significant effect on the differences, investigating the possible causes. Finally, according to our findings, we give some recommendations to web developers, browser vendors, and end users. 
\end{abstract}

\section{Introduction}
Web browsing is always the major requirement of Internet users. Recent reports show that the number of smartphone users worldwide has been over 2.6 billion, and the web traffic volume from mobile devices has exceeded that from desktop PCs. Compared to desktop PCs, web browsing on smartphones considers more on the quality of experiences (QoE): not only the look-and-feel and page loading time, but also the data traffic and energy consumption. Designing efficient mobile browser becomes a very significant strategy for browser vendors to grab the access channel of mobile web. Although it is argued that smartphone users prefer specific-purpose apps such as news reader and social network, more and more apps essentially integrate and customize browser engine for fetching and rendering content~\cite{3rdwebview}. For example, the BBC news app integrates Android WebView (which is a browser component based on Chrome) to exhibit its news list and news content.

Besides the major browsers such as Google Chrome, Mozilla Firefox, Opera, and Safari, it is observed that over 100 browsers developed by small-and-medium vendors have been released on popular app stores such as Apple AppStore, Google Play, and so forth. For example, in China, numerous browsers on smartphones are emerging, like UCWeb browser, QQ browser, 360 browser~\cite{browCNZZ}. Hence, smartphone users can have various alternatives to select browsers. Our previous study demonstrated that users holding different device models can have different preferences in choosing web browsers, i.e., the high-end users tend to choose Google Chrome while the low-end users tend to choose Opera and UC Web~\cite{Li:IMC15}. We can infer some possible reasons from the textual descriptions of browsers. For example, the UCWeb browser claims to employ cloud to compress the contents in advance before sending the data to the smartphones, and thus can save traffic for end-users who have very limited data plan~\cite{UCWeb}. Unfortunately, so far we have the absence of a systematic and comprehensive to demystify how these browsers differ from one another when fetching and rendering the same webpages. Two interesting but non-trivial questions need to be answered:
\begin{itemize}
\item How long do different mobile browsers load the same webpage?
\item How well does the cache performance vary in different browsers and thus affect data consumption for the same webpage accordingly?
\end{itemize}

Understanding the keys to these questions and their implications is vital on several aspects. With the increasing popularity and diversity of mobile devices for accessing the web, it is important for browser vendors to identify the aspects that can impact the quality of user experiences, so that they can fix potential defects and flaws and thus attract more users in the browser war. In addition, the mobile app developers can choose and integrate the adequate browser kernels into their apps for better browsing experiences. On the other hand, as web providers increasingly incorporate third-party services such as advertising, analytic, and Content Delivery Network (CDN) into their webpages, they need tools and techniques to evaluate the impacts of these services to users. Furthermore, beyond the perspective of any given users or web providers, understanding differences of mobile browsers is the first step toward solutions to webpage customization for varying client programs to achieve the right balance between performance, usability, and user interests.

This paper tries to bridge such a knowledge gap. We conduct a measurement study by employing three representatively popular mobile browsers: Chrome FireFox and Opera. We deploy the three browsers under the same smartphone emulator and made them request the same webpages under the same network environment. To avoid bias, our extensive measurement keeps on revisiting the homepages of 337 popular websites every 30 minutes, and continuously lasts for 3 weeks. To rigorously and comprehensively measure the differences among browsers, we devise a model with various metrics and apply statistical correlation analysis. The results demonstrate that a large percentile of webpages are significantly various on different browsers in terms of loading time and cache performance. Based on the findings, we suggest some implications to browser vendors, web developers, and end-users.

Although there have been numerous debates of browser war, to the best of our knowledge, we make the first study of comprehensively comparing the differences of various mobile browsers. More specifically, this paper makes the following contributions.

\begin{itemize}
\item We first report the diversity of choosing mobile browsers based on date usage logs collected from millions of smartphone users.
\item We establish an automated data-collection platform to collect fine-grained traces of the same webpages accessed from different mobile browsers, under the same OS, hardware, and network environment. We collect request traces of homepages from 337 popular websites continuously for three week.
\item We use various metrics to comprehensively characterize the differences among different browsers We reveal the statistically significant differences among representatively popular browsers in terms of loading time and data consumption when visiting the same webpage.
\item We analyze the root causes leading to the differences among browsers. For example, some kernel-related declarations in the CSS sheet, like \texttt{-webkit} \texttt{-text-sizeadjust}, \texttt{-webkit-appearance}, lead to special display effect in a certain browser. We then present some recommendations on how to improve the design of browsers and webpages. Meanwhile, end-users or app developers can take knowledge away from our findings to select the proper browser (or kernels) when browsing/designing specific webpages.
\end{itemize}

%

The remainder of this paper is organized as follows. Section~\ref{Motivation} evidences that current smartphone users tend to install various mobile browsers from a collection of 5-month logs of over 4 million real-world Android users. Section~\ref{Methodology} introduces our measurement methodology and the setup of our measurement platform. Section~\ref{data} describes our collected data. Section~\ref{loadtime} and \ref{cache} compare the loading time and cache performance of the same webpage on different browsers, respectively. Section~\ref{implication} provides some suggestions and recommendations to browser vendors, web developers, and end-users. Section~\ref{relatedwork} presents the related work that has been made in mobile web browsing. Section~\ref{conclusion} ends up the paper with concluding remarks and future work.


\section{Motivation}
\label{Motivation}
There are various kinds of browsers nowadays for mobile devices. Different browsers have different kernels, data persistence mechanisms and user interfaces. In this section, we use smartphone usage logs collected from millions of real-world Android users to present two interesting findings that illustrate the usage of mobile browsers. The data is from Wandoujia\footnote{Visit its official site via \url{http://www.wandoujia.com}.}, a free Android app marketplace in China. More details of the Wandoujia dataset can be referred in our previous work~\cite{Li:IMC15}\cite{TSE17Liu}\cite{TOIS17Liu}, including network activity statistics, permission monitoring, content recommendation, etc. We filter out a subset of the whole dataset related to only those browser apps. The filtered dataset contains 4-month records of browser app installation from 4,775,536 users.

\subsection{Diverse Choice of Browsers}
We first demonstrate the diverse choice of browsers on smartphones by considering the screen size. For each of top 10 mobile browsers, we calculate the cumulative distribution function (CDF) of devices with different screen sizes that has install the browser.

\begin{figure}[!t]
\centering
\includegraphics[width=0.45\textwidth]{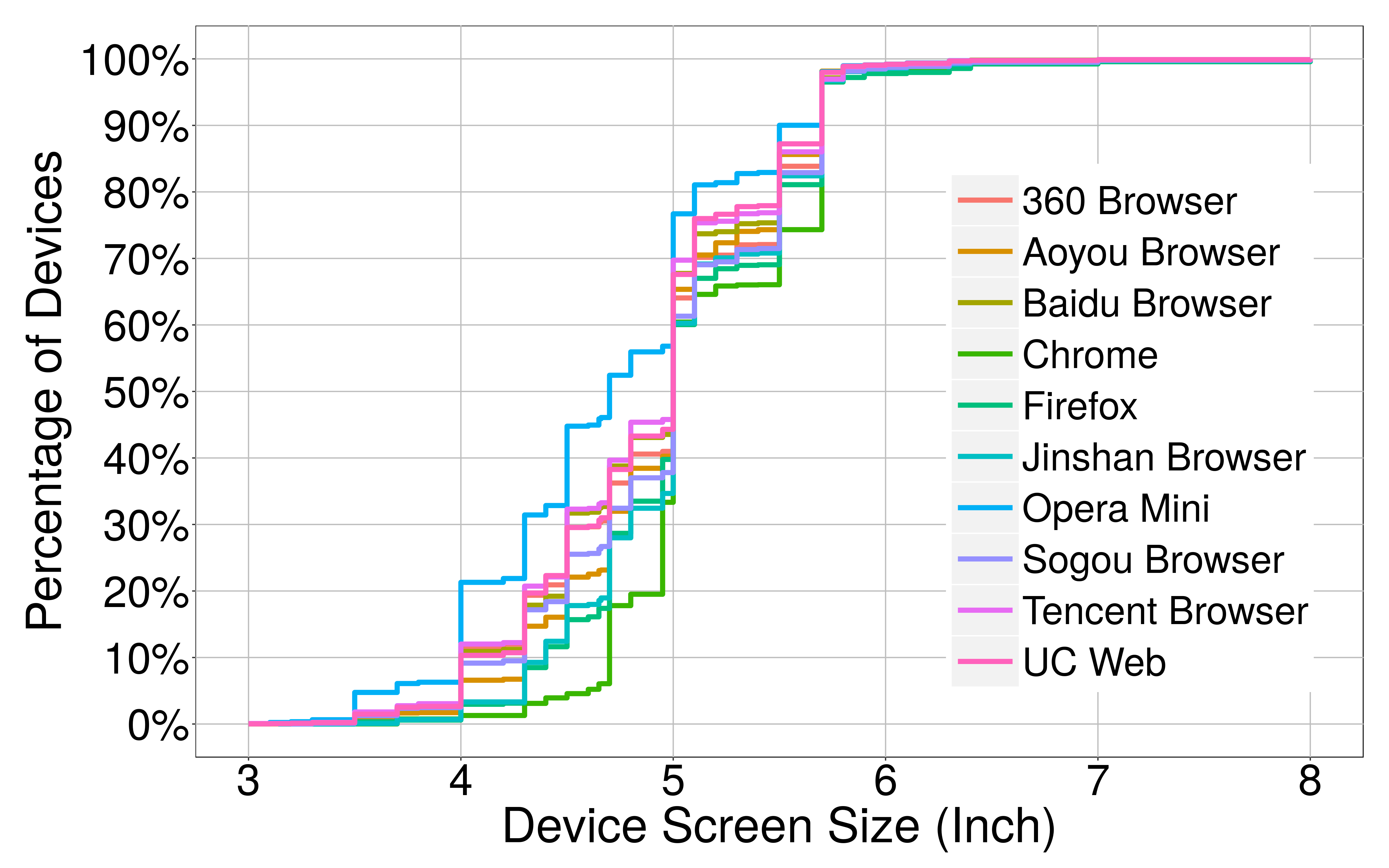}
\caption{Diversity of browser choice on smartphones}\label{fig:size}
\end{figure}

Figure~\ref{fig:size} depicts the result. It is easy to find that the screen size of mobile devices has a impact on the user selection of browsers. The line of Opera Mini is to the most upper-left, meaning that users of Opera Mini tend to use devices of smaller screen. On the contrary, the line of Chrome is to the most bottom-right, meaning that Chrome tends to be installed on devices of larger screen.

The finding reveals the subtle differences among browsers. We speculate that it can be caused by discrepant display of browsers, which precipitate users to choose browsers based on their screen size.

\subsection{Usage of Multiple Browsers}
\begin{figure}[!t]
  \centering
    \includegraphics[width=0.45\textwidth]{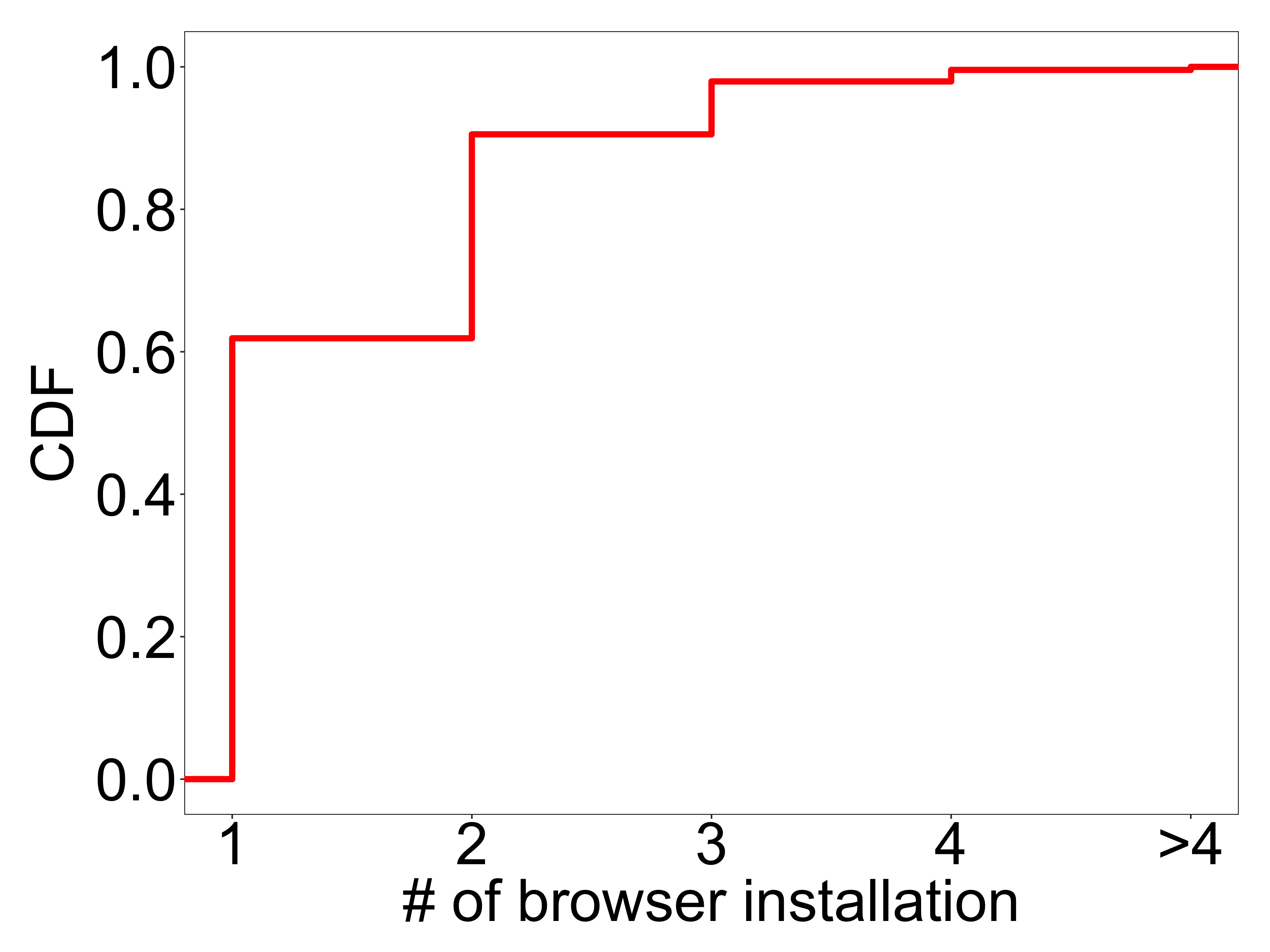}
  \caption{CDF of users who have installed multiple browsers. The x-axis is the number of browser installation. The y-axis is the CDF of users.}\label{figure:multipleusage}
\end{figure}

We then calculate how many browsers a single user has installed on his/her smartphone. Figure~\ref{figure:multipleusage} shows the result. About 38\% of users install more than one browser. There are even 1\% of users installing four or more browsers.

It is interesting to explore reasons why users need multiple browsers on their mobile devices. A clear explanation may be that the functions of a webpage may vary across different browsers. The browser compatibility issue has always been argued since the birth of the browser. However, apart from functions, mobile browsing considers more on non-functional issues such as the time spent on loading a page, and network traffic used to load a webpage.

\section{Methodology}
\label{Methodology}
In this section, we present how we study the quality of experience (QoE) of different mobile browsers.

Our study focus on the non-functional aspects of QoE, which is orthogonal to those of browser compatibilities.  We assume that the root cause to the QoE differences lies in the variety of resources acquired via different browsers. Based on the assumption, we design and establish a platform to collect traces of resources from 337 popular websites, via three popular mobile browsers, and for a considerably long time. The collected data form the basis of our measurement study.
\subsection{QoE Measurement}
For web browsing on mobile devices, QoE not only focuses on the page load time like browsing on desktop PCs, but also emphasizes the data and energy consumption since mobile devices usually access the Internet by cellular data plan and are battery-powered. Therefore, we focus on the \textbf{loading time} and \textbf{cache performance} of different browsers in this paper because. While loading time directly influence the user-perceived QoE, cache performance determines the data consumption. Energy consumption is not easy to compare among browsers but it can be induced from the loading time and cache performance.

Basically, there are many factors that could influence the QoE of different browsers, such as how well the browser's engine perform, and whether web developers optimize their websites for different browsers. However, the process of browsing the web on a browser is actually fetching a series of resources from the network. Therefore, we assume that the fetched resources when browsing the web on different browsers is a good indicator to the QoE differences when studying the root causes related to different browsers.
\subsection{Data Collection}
To analyze the QoE of different browsers, we need to collect real access data. However, real user data is not comparable due to the individual browsing behaviours. In addition, since we focus on only the browsers, we should control the hardware, OS and network environment, and visit same webpages simultaneously on different browsers.
\subsubsection{Webpages}
We select the webpages to be studied from homepages of popular websites ranked by Alexa~\cite{Alexa} top 500 list. For some problems, we filtered out some of the following webpages:
\begin{itemize}
\item	Unreachable webpages. Some webpages are not available because of server internal error or network problem.
\item	Duplicate webpages. Some webpages have different domain names but actually offer the same service. We keep only one of them.
\end{itemize}

Finally, we have a set of 337 webpages.
\subsubsection{Design of data-collection platform}
\begin{figure}
\centering
\includegraphics[width=0.5\textwidth]{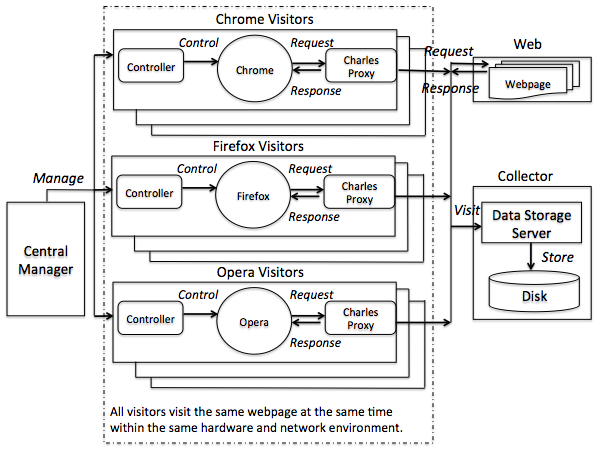}
\caption{The architecture of data-collection platform.}\label{fig:platform}
\end{figure}

We design a data-collection platform that can control different browsers to simultaneously visit same webpages and record the network traffic during the process. Figure~\ref{fig:platform} shows the architecture of the platform.

A visitor is a node that requests a set of webpage at a certain interval. The webpages are identified by a list of given URLs. A visitor is composed of three parts: a browser, an extension and a Charles web proxy~\cite{Charles}. We enable browsers' emulation mode and set the user agent to ``Android 4.2". In this case, the servers will return mobile-version webpages. When browsers render the webpage, resources are requested either from the network or from the local cache. To study the cache performance, in some nodes, the cache policy is set to be prohibited in browser configuration in order to acquire the whole set of resources at each time point. In other nodes, we enable the cache to study the real browser behaviour.

The Charles web proxy is used to record all network traffic. Charles records all HTTP/HTTPS connections passing through it. Moreover, Charles offers a set of APIs for programmers to access its data, which helps us download and save.

Finally, we build extensions for browsers to automatically revisit a webpage. The extensions are a little bit different from each other because different browsers offer different interfaces for extensions. However, the control logics remain the same: all extensions are set to revisit a list of given URLs at a certain interval, then access data recorded in Charles and submit the response to the collector. To make sure that a webpage has been loaded completely before shifting to the next one, we set a threshold of one minute. That's to say, the next webpage will wait for at least one minute after the previous webpage being requested.

The collector is a PHP server receiving all data collected by visitors. The PHP server receives data from all visitors and record information to its local disk.

The central manager is used to control all the visitors. It sets the starting time of data collection in order to make all browsers start their work simultaneously. It can also stop or restart a certain browser to handle some exceptions during the collection.

\subsubsection{Platform deployment}

\begin{table}[t]
\centering
\caption{Deployment environment}\label{table:platform}
\begin{tabular}{|p{2.9cm}|p{5.5cm}|}
\hline
\multirow{3}{2.9cm}{Browser Version} & Chrome 41.0.2272\\
\cline{2-2}
&Firefox 36.0\\
\cline{2-2}
&Opera 28\\
\hline
Hardware & 1G Memory, 1 core CPU\\
\hline
Operating System & Ubuntu 14.04 LTS\\
\hline
Network & 1M bandwith Ethernet of Ali Cloud\\
\hline
Location & Qingdao, China\\
\hline
\end{tabular}
\end{table}

We deploy our platform on Aliyun virtual machines and make sure that all visitors visit the same webpage at the same time within the same hardware and network environment. Table~\ref{table:platform} shows the deployment details. We choose three most popular mobile browsers: Chrome, Firefox and Opera, as the browsers to investigate. All the virtual machines are located in the same geographical location in Qingdao, China to avoid the CDN's influence.

We carried out 24 groups of data collection at the same time: In 3 groups, we use Chrome, Firefox or Opera, to revisit a list of mobile webpages every 30 minutes, with the browser cache disabled. In the other 21 groups, the browser cache is enabled, and the revisiting intervals are 30 minutes, 6 hours, 12 hours, 1 day, 2 days, 4 days and 7 days. We control to make sure that all these 24 groups visit a certain webpage simultaneously. We keep track of all request and response traffic for three weeks. That is approximately 1000 visits for each webpage. The total size is 4.6 terabytes. After parsing saved files, we have a database of over ten million records.

\section{Data Characterization}
\label{data}
In this section, we first describe the data set of our experiment. Then we describe the factors and metrics we use to illustrate the webpage and explain the QoE difference on different browsers.

\subsection{Factors and Metrics}
We consider a list of factors that may be correlated with the QoE. In later section, we analyse the impacts of these factors and illustrate the major factors that cause QoE differences on different browsers. These metrics describe all the factors of the content of a webpage. Table~\ref{table:factor} is the list the of metrics.
\begin{table*}[t]
\footnotesize
\centering
\caption{Metrics definition}\label{table:factor}
\begin{tabular}{|p{3cm}|p{5cm}|p{5cm}|p{3cm}|}
\hline
Factors & Metrics & Explanation & Rationale\\
\hline
\multirow{2}{2.9cm}{Scale of resources}& Numbers of resources (resources number) & Total numbers of resources in a webpage. & \multirow{2}{2.9cm}{Have impacts on loading time and cache performance}\\
\cline{2-3}
&Size of resources (resources size) &Total size of resources in a webpage. (Measured by Byte)&\\
\hline\end{tabular}

\begin{tabular}{|p{3cm}|p{5cm}|p{5cm}|p{3cm}|}
\hline
\multirow{5}{2.9cm}{Distribution of MIME type resources}& Percentage of JavaScript resources (js proportion) & Mean proportion of JavaScript resources in a webpage. & \multirow{5}{2.9cm}{Have impacts on loading time and cache performance}\\
\cline{2-3}
&Percentage of HTML resources (html proportion) &Mean proportion of HTML resources in a webpage&\\
\cline{2-3}
&Percentage of style sheet resources (css proportion) &Mean proportion of css and xslt resources in a webpage&\\
\cline{2-3}
& Percentage of image resources (img proportion) &Mean proportion of image resources in a webpage&\\
\cline{2-3}
&Percentage of json resources (json proportion) &Mean proportion of json resources in a webpage&\\
\hline\end{tabular}
\begin{tabular}{|p{3cm}|p{5cm}|p{5cm}|p{3cm}|}
\hline
\multirow{3}{2.9cm}{Dependence on third party webpages}& Percentage of third party resources (thirdparty proportion) & Mean proportion of third party resources in a webpage. & \multirow{3}{2.9cm}{Have impacts on loading time}\\
\cline{2-3}
&Percentage of browser-offering resources (browser proportion) &Mean proportion of third party resources offered by browser in a webpage&\\
\cline{2-3}
&Percentage of ads resources (browser proportion) &Mean proportion of ads in a webpage&\\
\hline\end{tabular}

\begin{tabular}{|p{3cm}|p{5cm}|p{5cm}|p{3cm}|}
\hline
HTTPS secure mechanism&Percentage of https resources (https proportion) & Mean proportion of resources using https protocol. &Have impacts on loading time\\
\hline\end{tabular}

\begin{tabular}{|p{3cm}|p{5cm}|p{5cm}|p{3cm}|}
\hline
Redundant transmission& Numbers of invalid resources (invalid number) & Total numbers of invalid resources when requesting for a webpage. & Have impacts on cache performance\\
\hline\end{tabular}

\begin{tabular}{|p{3cm}|p{5cm}|p{5cm}|p{3cm}|}
\hline
Explicit cache control& Percentage of cache explicitly set resources (cached proportion) & Mean proportion of resources with cache policy explicitly set.& Have impacts on cache performance.\\
\hline\end{tabular}

\end{table*}

Firstly, we attempt to define several factors that potentially influence the loading time of a webpage.

\textbf{Scale of resources} of a webpage includes the total number of resources and total size of resources. The resources in a webpage include all HTML, JavaScrpt, CSS, image, etc. after being rendered. The scale of resources can vary a lot among different browsers. A Browser has a unique identifier in the \texttt{user-agent} field of an HTTP request to identify its type and version. For example, requests sent by Chrome include \emph{Chrome /41.0.2272} in user-agent field. While requests sent by Firefox include \emph{Gecko/20100101 Firefox/36.0} in user-agent field. Thus, the servers may return different resources to different browsers, leading to different sizes and numbers of resources. Scale of resources lead to different loading time on different browsers because large number of resources tend to consume much more time. What's more, browsers may have their own strategies to parse and render a webpage. For example, the pre-fetching strategy may be suitable for webpages with large numbers of resources, because it saves the time for HTTP connection.

\textbf{Distribution of MIME type resources} messures the distribution of HTML, JavaScrpt, CSS, image and other resources in a webpage. It includes the size and number of HTML, JavaScrpt, CSS, image and other resources in a webpage. We concern about this factor because browsers may show different capabilities of handling these resources. For example, a browser that has a faster JavaScript engine may have less loading time for webpages with a high proportion of JavaScript resources.

\textbf{Dependence on third-party resources} influences the loading time among browsers. We judge whether a resource is from a third-party website by its URL. If the URL contains the key word of a website's domain, then it is identified as the normal resource provided by the web developer. Otherwise, it is identified as a third-party resource provided by other developers. What's more, we are also interested in two kinds of third-party resources. The first one is the ads. The second one is the resources offered by the browser vender. We list the numbers of these resources, too. Browsers may show different preferences to third-party resources and even forbid some of the domains due to security reasons.

\textbf{HTTPS secure mechanism} offers extra security guarantee for users. However, HTTPS requires bidirectional encryption of communication and the capability of different browsers may be vary. The numbers of HTTPS resources is also what we should concern about.

Then, we define several factors that potentially influence the cache performance.

\textbf{Scale of resources} has an impact on the total data traffic in one visit. However, in later analysis, we tend to use the proportion to the total traffic, which reduces the impact of this factor.

\textbf{Distribution of MIME type resources} not only has an impact on loading time, but also influences the data cosumption. Some specific types of resources, like JavaScript, are much more variable than the others. High percentage of those kinds of resources leads to more data consumption.

\textbf{Redundant transmission} consists of invalid resources with unexpected HTTP status code and duplicated resources. E.g., a resource responding with status 404 is invalid, and a resource having totally same URL with another one is duplicated. Redundant transmission may be caused by poor strategy of error handling. It is one of the reasons that cause more data consumption.

\textbf{Explicit cache control of resources} is an important factor we should concern. Cache policy influences the data consumption in an obvious ways. A webpage with larger percentage of resources that have explicitly set cache parameters is likely to consume less data in the long term.

\subsection{Difference of Metrics on Different Browsers}
Now we compare the resources of webpages using the factors and metrics we have defined. For each of these 337 webpages, we have a trace of over 1000 visits using Chrome, Firefox and Opera. We first analyze the statistical significance of the difference between two browsers, i.e., Chrome vs. Firefox, Firefox vs. Opera, Chrome vs. Opera, by applying Mann-Whitney U test at p-value=0.05. E.g., when we compare the total resource number of different browsers, we need to carry out three U tests. The first null hypothesis is that Chrome does not have significant different numbers of resources in a visit with Firefox, the second one is that Firefox does not have significant different numbers of resources in a visit with Opera and the third one is that Chrome does not have significant different numbers of resources in a visit with Opera. Thus, we can see whether a webpage have different total numbers of resources on any two of the browsers.

\begin{table}[t!]
\footnotesize
\centering
\caption{Percentage of browsers with significant difference between two browsers measured by a certain metric.}\label{table:common}
\begin{tabular}{|p{1.75cm}|p{1.75cm}|p{1.75cm}|p{1.75cm}|}
\hline
Metrics &\small Percentage of webpages with significant difference between Chrome and Firefox &\small Percentage of webpages with significant difference between Firefox and Opera &\small Percentage of webpages with significant difference between Chrome and Opera \\
\hline
total number & 91.39\% & 94.65\% & 84.56\% \\
\hline
total length & 90.16\% & 92.02\% & 82.09\% \\
\hline
html proportion & 58.46\% & 97.03\% & 97.03\% \\
\hline
js proportion & 60.83\% & 61.42\% & 33.53\% \\
\hline
css number & 33.23\% & 34.72\% & 18.99\% \\
\hline
img proportion & 76.85\% & 74.48\% & 50.44\% \\
\hline
json proportion & 22.55\% & 23.03\% & 12.76\% \\
\hline
thirdpart number & 87.53\% & 93.77\% & 82.79\% \\
\hline
ads number & 31.75\% & 32.94\% & 21.66\% \\
\hline
browser number & 81.60\% & 98.81\% & 94.66\% \\
\hline
redundant number & 64.99\% & 65.58\% &35.61\% \\
\hline
cached number & 79.23\% & 93.18\% & 89.91\% \\
\hline
https number & 72.70\% & 56.97\% & 64.98\% \\
\hline\end{tabular}
\end{table}
Table~\ref{table:common} lists the percentage of webpages that show significant difference after U test for each metric. It is amazing that, most metrics, like the total scale of the website (measured by total number and total size), the distribution of different types of resources (measured by different MIME type resources) and dependence on third-party resources, show great difference among browsers. For example, up to 90\% webpages have different scales among Chrome and Firefox.

\subsection{Common Resources of Browsers}
To have a better understanding of the differences, we classify all the resources into seven categories: the resources shared by all browsers (ALL), the resources shared by Chrome and Firefox but not Opera (CRFF), the resources shared by Firefox and Opera but not Chrome (FFOP), the resources shared by Opera and Chrome but not Firefox (OPCR), the resources fetched only by Chrome (CR), the resources fetched only by Firefox (FF) and the resources fetched only by Opera (OP).

Now we take the comparison between the resources of these seven categories. We calculate the proportion of different kinds of resources.

\begin{table}[ht]
\centering
\caption{The MIME type distribution in seven categories.}
\begin{tabular}{|p{0.9cm}|p{0.9cm}|p{0.9cm}|p{0.9cm}|p{0.9cm}|p{0.9cm}|p{0.9cm}|}
\hline
\tiny{Category} & \tiny{HTML}& \tiny{JavaScript} &\tiny{Image}&\tiny{CSS}&\tiny{JSON}&\tiny{Others}\\
\hline
CR & 41.52\% & 29.77\% & 13.68\% & 1.59\% & 4.58\% & 8.86\%\\
\hline
FF & 39.27\% & 28.92\% & 17.13\% & 1.88\% & 4.21\% & 8.59\%\\
\hline
OP & 49.65\% & 29.40\% & 12.60\% & 1.17\% & 4.62\% & 2.56\%\\
\hline
CRFF & 21.53\% & 8.56\% & 34.09\% & 2.86\% & 3.13\% & 29.83\%\\
\hline
FFOP & 22.33\% & 9.17\% & 33.00\% & 1.59\% & 4.58\% & 29.33\%\\
\hline
OPCR & 23.16\% & 11.68\% & 37.79\% & 2.48\% & 3.50\% & 21.39\%\\
\hline
ALL & 14.59\% & 21.33\% & 50.07\% & 6.76\% & 2.35\% & 4.90\%\\
\hline\end{tabular}
\end{table}

The common resources of all three browsers have a high proportion of image resources while have a low proportion of HTML resources. The distinct resources fetched by only one browser have a high proportion of HTML while have a low proportion of image. The results show that within all kinds of resources, HTML tends to be different among browsers while image tends to be the same.

\subsection{Examples of Differences}
To make it more intuitive and comprehensible, we list several common differences between the content of two browsers at a simultaneous visit (shown in Figure~\ref{figure:exmaple}.

\begin{figure*}[!t]
\centering
	\subfigure[\emph{Espn.go.com} on Chrome]{
    \label{fig:realworld:a} 
    \includegraphics[width=0.32\textwidth]{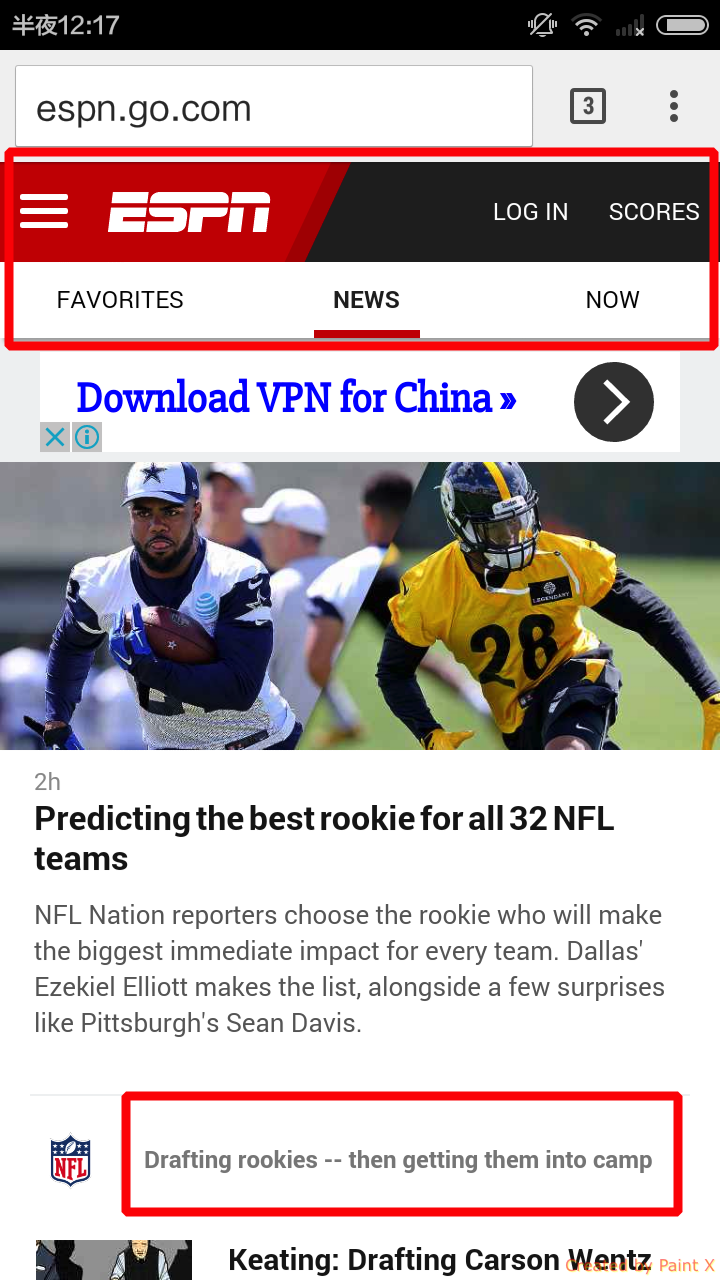}}
    \subfigure[\emph{Espn.go.com} on Firefox]{
     \label{fig:realworld:a} 
    \includegraphics[width=0.32\textwidth]{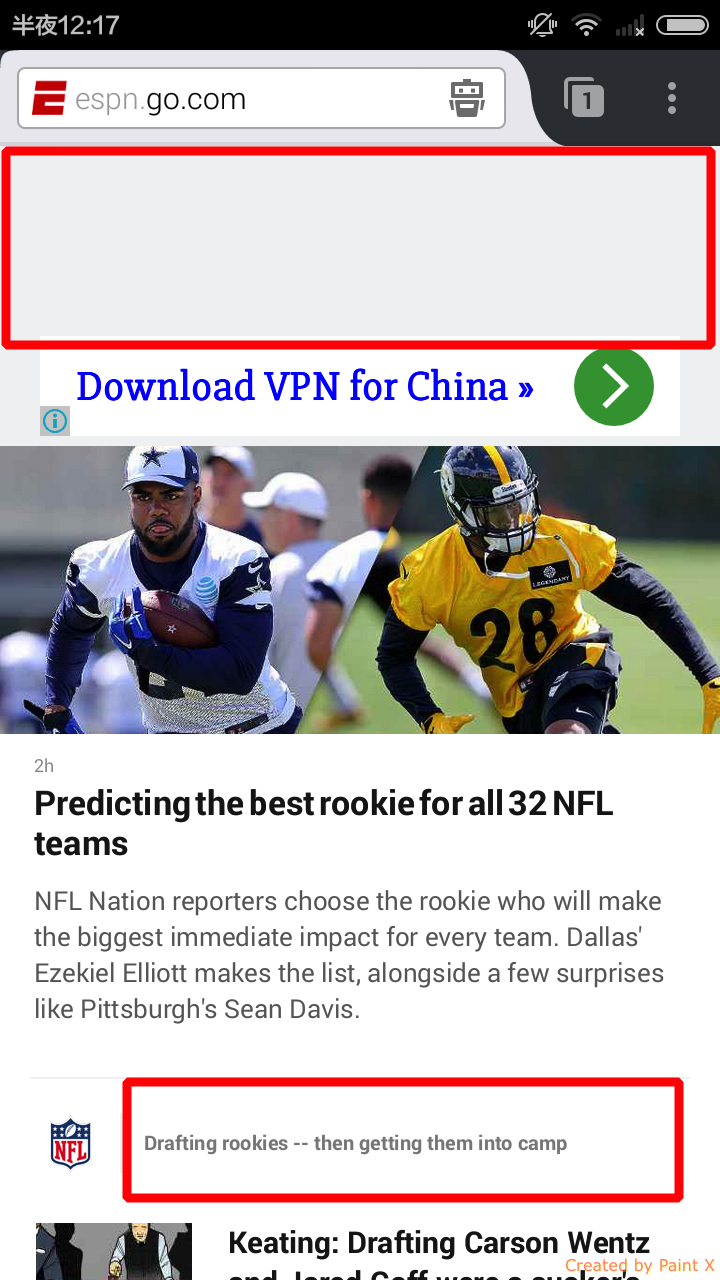}}
    \subfigure[\emph{Espn.go.com} on Opera]{
     \label{fig:realworld:a} 
    \includegraphics[width=0.32\textwidth]{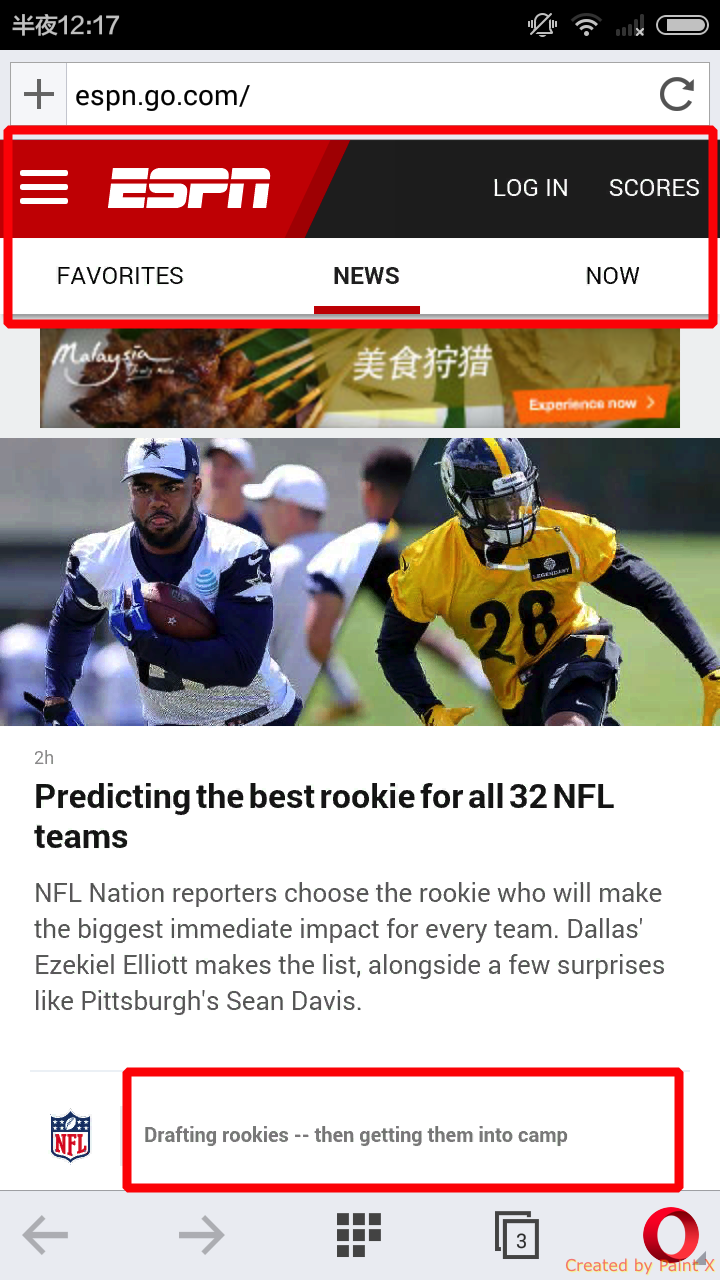}}
\caption{Website \emph{Espn.go.com} on different browsers. Differences are shown in red box. Firefox don't show tool bar on the top. While Chrome has larger font size than the others.}\label{figure:example}
\end{figure*}

\textbf{Numbers of resources} in a webpage can be different. For example, when visiting \emph{Espn.go.com}, Firefox is lack of lots of resources and doesn't show tool bar on the top.

\textbf{Ads} make differences among browsers. In some cases, different browsers recommend different ads for users. In some cases, browsers don't show ads at the same time or at the same position on the screen.


\textbf{Kernel related style sheet} can be different between two browsers. In the visiting traces of \emph{baidu.com}, we find that the document of Opera contains a declaration of extra css style, while Firefox doesn't. The reason is that this style declaration is related to the browser kernel. To improve readability on webpages designed for desktop browsers, mobile browsers with webkit kernel automatically increases the size of small text. For Firefox, it's non-standard. There are several attributes in this category: -webkit-text-sizeadjust, -webkit-appearance, -webkit-box-sizing and etc.
\begin{lstlisting}[language={C},keywordstyle=\color{blue!70},commentstyle=\color{red!50!green!50!blue!50},frame=shadowbox, rulesepcolor=\color{red!20!green!20!blue!20},basicstyle=\small\ttfamily]
<style type="text/css" class="spa-index-style">
body {
	font-family:Arial,Helvetica,sans-serif;
	-webkit-text-sizeadjust:none;}
input,button,textarea {
	-webkit-appearance:none;
	-webkit-box-sizing:border-box;}

...
</style>
\end{lstlisting}
The special adjustment of style sheet is necessary. There are several other differences about the style sheet. E.g., not all browser engines are allowed to control the final size of the text using a percentage value (Webkit and Trident do allow it, Gecko doesn't). Thus, webpages using percentage value need different style sheets for different browsers.

\textbf{Extra webpages provided by browser vendors} is a common difference. E.g, Opera provide a so called fraud check for current opening site. For any opened site, the browser sends a request to a page( \emph{sitecheck2. opera.com} ) with the current link as the parameter. Consequently, all webpages rendered on Opera have this special request. Not only Opera, but also vendors of Chrome and Firefox provide their special services to the users, too. Extra webpages provided by browser vendors are totally different to users.

\section{Loading Time Comparison}
\label{loadtime}
Loading time refers to the time before a website completely loaded and being capable of reacting to the next request of users. It is one of the major targets to improve the quality of user experience. Loading time doesn't include the propagation delays; because web providers and web users are distributed all around the world and the propagation delays may be vary. To measure the loading time of different websites, we focus on the time of resolving, analyzing rendering and display. In this section, we aim to compare the loading time of the same website on different browsers.

\subsection{Comparison of Loading Time}
\begin{figure}[!t]
\centering
\includegraphics[height=2.2in, width=3in]{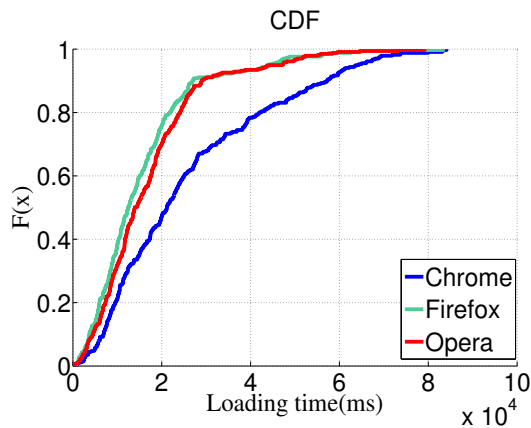}
\caption{The cumulative distribution function of loading time on different browsers. The x-axis is the loading time. The y-axis is the CDF.}
\end{figure}
After gathering a set of 337 websites` traces, we now have to measure the loading time of these websites on different browsers. Loading time is an important factor impacting human expectations, feelings, and satisfaction with respect to a particular product. We define the loading time as the time between the request time of the first resources and the response time of the last one. What's more, loading time of a website is defined as the median loading time in all visits

The cumulative distribution function of loading time on different browsers show explicit difference between each other. We are surprising to find that, the actual loading time of Chrome is larger than the other two. There are also some difference between Firefox and Opera.

\subsection{Website Classification}
Generally speaking, there is a great deal of difference on different browsers when we consider the loading time of websites on different browsers. However, When we consider a certain website, the loading time of this website on these three browsers depend not only on the features on browsers, but also the features of the website itself. We are interested about the features of those websites that show better loading performance on a certain browser.

For each of these 337 webpages, we have a trace of over 1000 visits using Chrome, Firefox and Opera. We can judge whether a website has significant less loading time compared to the other two browsers. We first analyze the statistical significance of the difference between two browsers, i.e., Chrome vs. Firefox, Firefox vs. Opera, Chrome vs. Opera, by applying Mann-Whitney U test at p-value=0.05. E.g., when we compare Chrome vs. Firefox, we need to carry out two single side tests. The first null hypothesis is that Chrome doesn't show advantage over Firefox and the second one is that Firefox doesn't show advantage over Chrome. These two statistical tests are carried to justify that on which browser a webpage has less loading time.

\begin{table}[ht]
\centering
\caption{Numbers and Percentages of websites which show significant disadvantages when compared to another.}
\begin{tabular}{|p{1.8cm}|p{1.2cm}|p{1.2cm}|p{1.2cm}|p{1.2cm}|}
\hline
\multirow{2}{1.5cm}{Browsers }& \multicolumn{2}{|c|}{Results}& \multicolumn{2}{|c|}{ Results}\\
\cline{2-5}
&\small{Numbers}&\small{Proportion}&\small{Numbers}&\small{Proportion}\\
\hline
Chrome vs. Firefox & 103 & 30.84\% & 120 & 35.93\% \\
\hline
Chrome vs. Opera & 101 & 30.24\% & 54 & 16.17\% \\
\hline
Firefox vs. Opera & 151 & 45.21\% & 95 & 28.44\% \\
\hline\end{tabular}
\end{table}

The table shows the numbers and percentage of webpages that are significantly different on different browsers measured by loading time. Statistical results of significant advantages mean that the first browser has significantly shorter loading time. Statistical results of significant disadvantages mean that the first browser has significantly larger or longer loading time. E.g., the first row of table shows that 30.84\% of webpages have smaller loading time on Chrome and 35.93\% of webpages have smaller loading time on Firefox.

We find that a large quantity of the selected webpages has statistically significant differences (about 46\% to 80\%). Webpages on Chrome are much more similar to Opera (46.41\% measured by data traffic) than Firefox (66.77\% measured by data traffic).

Then we have a closer look at each website and investigate each dimension one by one. We define three sets here:
\begin{itemize}
\item CR: Webpages have less loading time on Chrome
\item FF: Webpages have less loading time on Firefox
\item OP: Webpages have less loading time on Opera
\end{itemize}
A website is regarded as having less loading time on a certain browser when it significantly has shorter loading time than the other two browsers within all visits.

\begin{table}
\centering
\caption{Classification of websites based on their loading time on different browsers}
\begin{tabular}{|c|c|l|}
\hline
Set & Number & Proportion\\
\hline
CR & 66 & 19.58\% \\
\hline
FF & 95 & 28.19\% \\
\hline
OP & 53 & 15.73\% \\
\hline\end{tabular}
\end{table}

The numbers and proportions of each group are shown above. We are even more interested about another question: what kind of the webpages is suitable for a browser? In another word, why a webpage has significantly less loading time on a browser? We will answer this question in the next part.

\subsection{Relationship between Metrics and Optimal Browser}
Prior work has defined several metrics that describe the resource content on different browsers. In this section, we are interested in investigating how each factor of a website is related with its optimal browser.

We have defined 3 sets, i.e., CR, FF and OP to represent the set of webpages that are particularly suitable for Chrome, Firefox and Opera. We now compare the values of each factor between CR and CR`s complement, FF and FF`s complement, OP and OP`s complement.

We first analyze the statistical significance of the difference between the two sets by applying the U test at p-value=0.05. The null hypothesis is that webpages perform better on Chrome doesn`t have significant different factors with the others. To show the effect size (ES) of the difference between the two sets, we compute Cliffs Delta (d). d is used to measure how often the values in one distribution are larger than the values in second distribution.

Tables below show the d value of all factors. Statistically significant difference should have large d value. We interpret the effect size values as small for d\textless 0.1, medium for 0.1\textless = d\textless =0.15 and large for d\textgreater = 0.15.

Several metrics that have large cliffs d value are listed as follows:

The webpage suitable for Chrome, i.e. CR, tends to have a large proportion of JavaScript resources. Its d value is 0.176 and is much more significant than Firefox and Opera. In another word, webpages that perform better on Chrome have a large proportion of JavaScript resources. It can be reasonable because a browser can improve the performance by accelerate the rendering and execution of JavaScript.

CR tends to have a lower percentage of image files while FF tends to have a higher one. It means that Chrome and Firefox may have different policies on image rendering. Typically, Firefox may be good at handling large quantities of images including png, jpeg and gif.
CR and OP tend to have large proportion of text files including html and xml. FF has the lower one to the opposite.  It reflects the ability difference of rendering DOM node of different browsers.

OP has a large percentage of third-party resources, which means Opera provides better approaches to third-party webpages.

FF has a small percentage of HTTPS resources while OP has a great one. HTTPS consists of communication over Hypertext Transfer Protocol within a connection encrypted by Transport Layer Security or its predecessor, Secure Sockets Layer. HTTPS requires bidirectional encryption of communications, which means can influence the performance of a webpage.

Some kinds of MIME type resources, specificly CSS and JSON, don't have much influence on the loading time among browsers.

\begin{table}
\centering
\caption{Statistic results}
\begin{tabular}{|p{2.5cm}|p{0.4cm}|c|p{0.4cm}|c|p{0.4cm}|l|}
\hline
\multirow{2}{*}{Metrics}& \multicolumn{2}{|c|}{$CR vs \neg{CR}$ }& \multicolumn{2}{|c|}{$FF vs \neg{FF}$}& \multicolumn{2}{|c|}{$OP vs \neg{OP}$}\\
\cline{2-7}
&Rel&ES&Rel&ES&Rel&ES\\
\hline
resources numbers& + & 0.096 & - & 0.106 & + & 0.022 \\
\hline
resources size& + & 0.075 & - & 0.097 & + & 0.035 \\
\hline
js proportion& - & 0.176 & - & 0.004 & + & 0.003 \\
\hline
html proportion& - & 0.155 & + & 0.107 & - & 0.167 \\
\hline
img proportion& + & 0.172 & - & 0.118 & + & 0.050 \\
\hline
css proportion& - & 0.026 & + & 0.014 & - & 0.001 \\
\hline
json proportion& - & 0.011 & - & 0.034 & - & 0.031 \\
\hline
thirdparty numbers& - & 0.070 & + & 0.088 & - & 0.191 \\
\hline
https numbers& - & 0.099 & + & 0.128 & - & 0.191\\
\hline\end{tabular}
\end{table}

\section{Cache Performance Comparison}
\label{cache}
In this section, we first measure the cache performance on different browsers. Then we take a deeper discussion about the reasons that lead to the difference on cache performance.
\subsection{Cache Performance Measurement}
We use the cache hit rate to denote the cache performance. The hit rate is defined as the division of saved traffic and the total traffic in this visit. As we have collected the traces of both cached and uncached revisiting records, we simply get the hit rate from their division. What's more, hit rate of a website is defined as the median hit rage in all visits.

We show the median hit rate of all websites below.The x-axis is the revisiting interval, i.e, 30 minutes, 6 hours, 12 hours, 1 day, 2 days, 4 days, 1 week. The y-axis is the median hit rate of 337 websites. An obvious conclusion is that the median hit rate decreases as the revisiting time increases. What is more interesting is that the cache performance of Firefox is poorer than Chrome and Opera in almost all intervals, while Chrome and Opera are alomost the same.

\begin{figure}
\centering
\includegraphics[height=2.2in, width=3in]{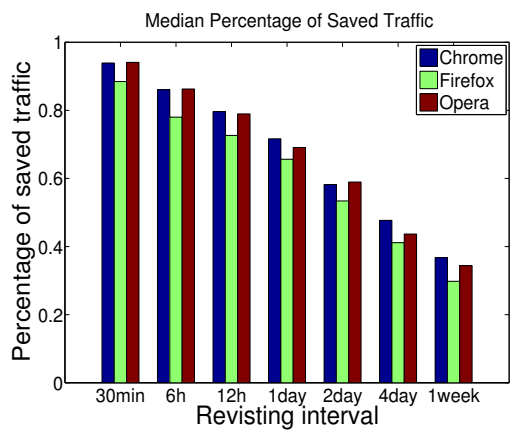}
\caption{Median hit rate of all websites in different revisiting interval. The x-axis is the revisiting interval, i.e, 30 minutes, 6 hours, 12 hours, 1 day, 2 days, 4 days, 1 week. The y-axis is the median hit rate of 337 websites.}\label{figure:cachetotal}
\end{figure}

We list the more detailed CDF plot of hit rate here. The x-axis is the median percentage of saved traffic of a website. The y-axis is the CDF. The revisiting interval from (a) to (g) is 30 minutes, 6 hours, 12 hours, 1 day, 2 days, 4 days, 1 week. The red line is Firefox, blue is Chrome and brown is Opera. It becomes more clear that the distribution of Firefox's hit rate is significantly smaller than the other two browsers. We try to find some causes to explain it.
\begin{figure*}[!ht]
  \centering
  \subfigure[]{
    \label{fig:realworld:a} 
    \includegraphics[width=0.32\textwidth]{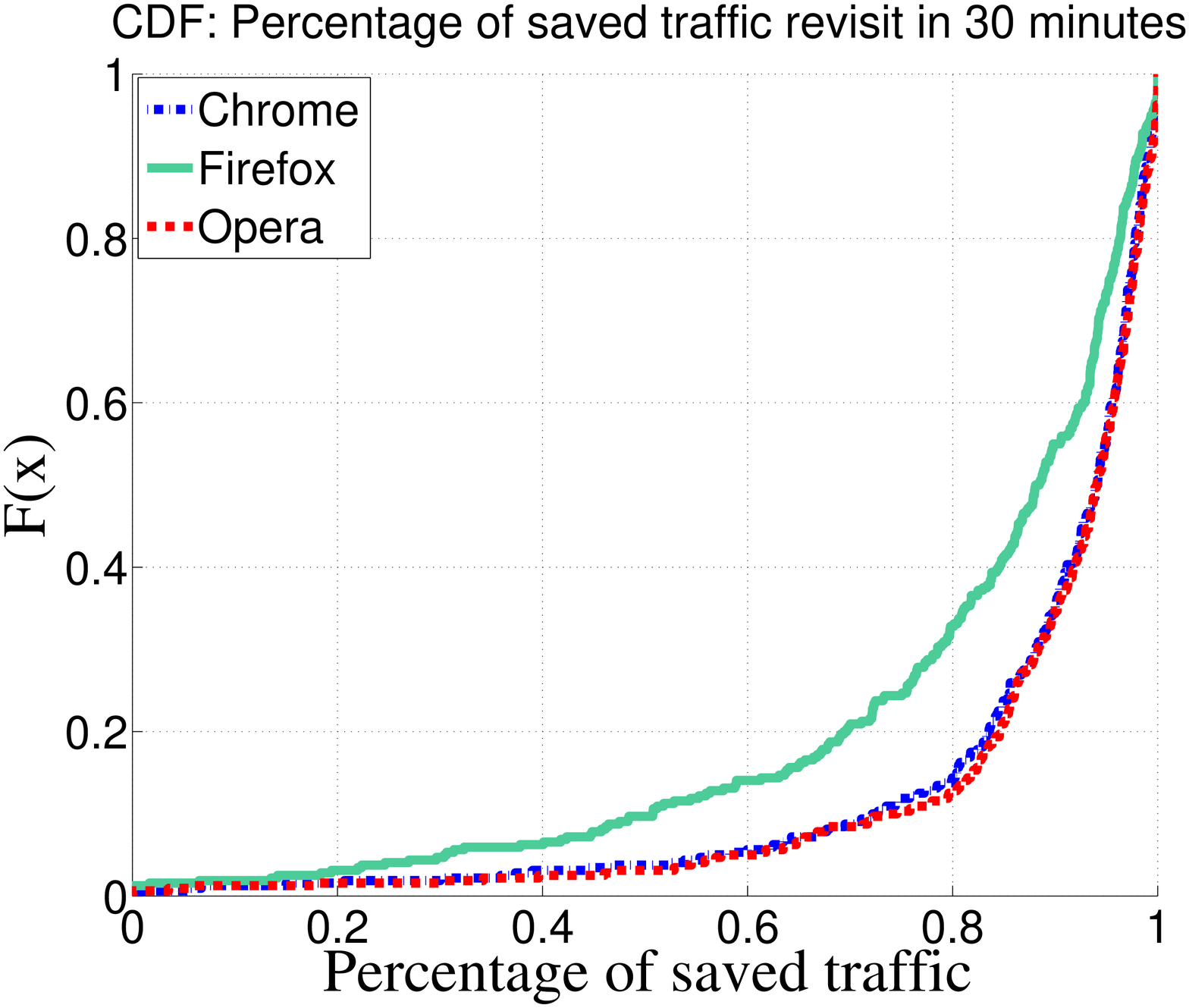}}
  \subfigure[]{
    \label{fig:realworld:b} 
    \includegraphics[width=0.32\textwidth]{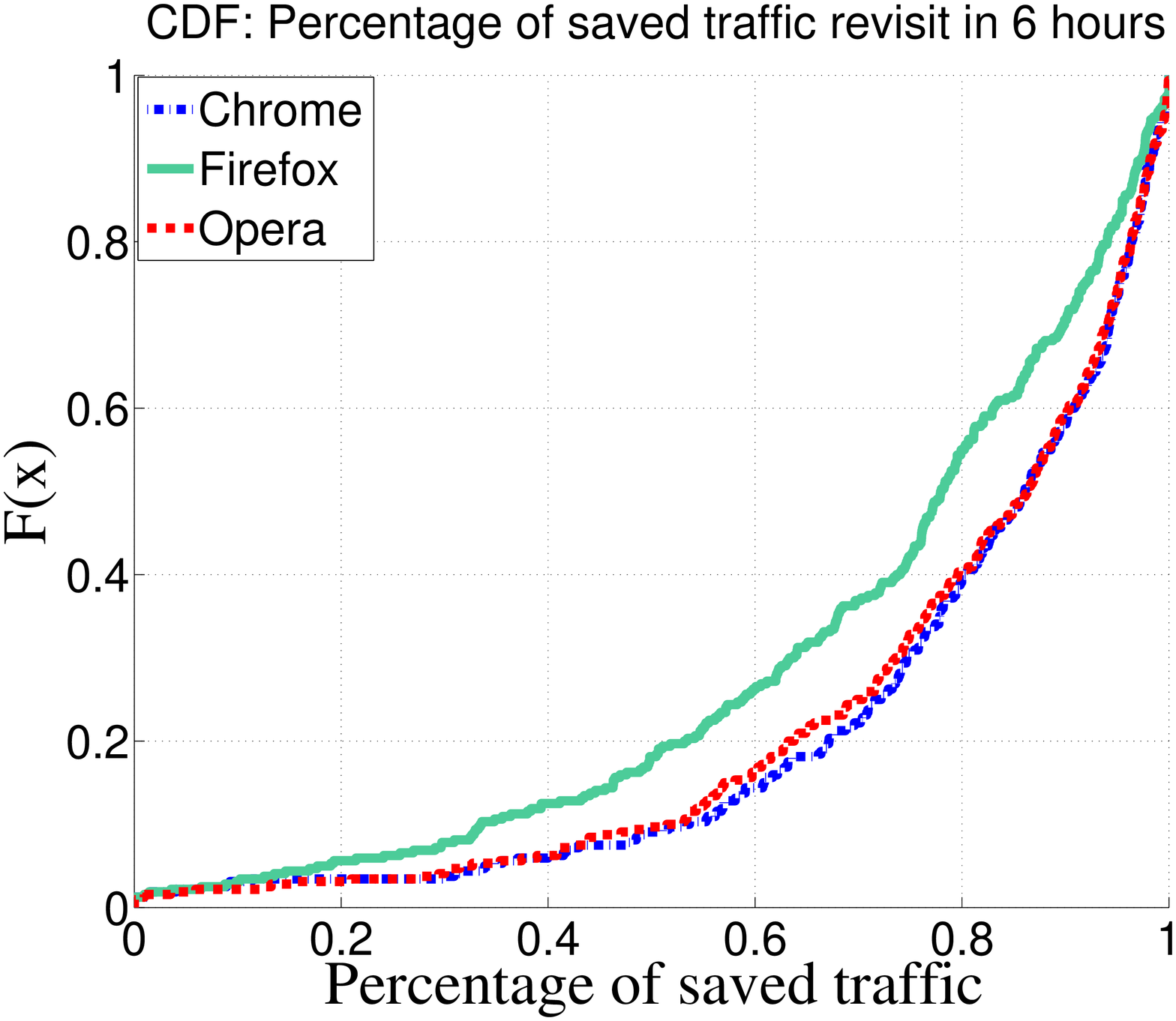}}
  \subfigure[]{
    \label{fig:realworld:c} 
    \includegraphics[width=0.32\textwidth]{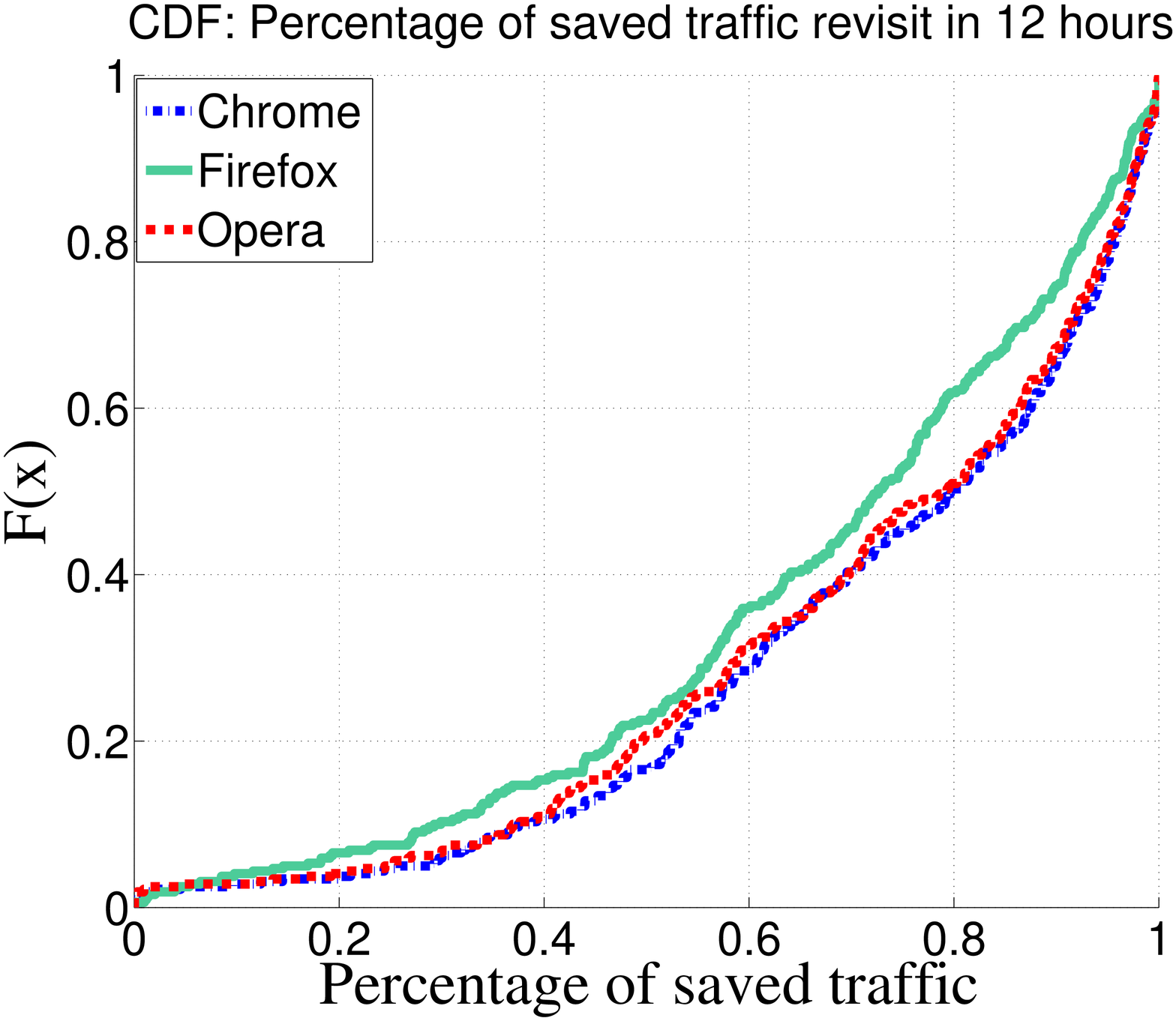}}
  \subfigure[]{
    \label{fig:realworld:d} 
    \includegraphics[width=0.32\textwidth]{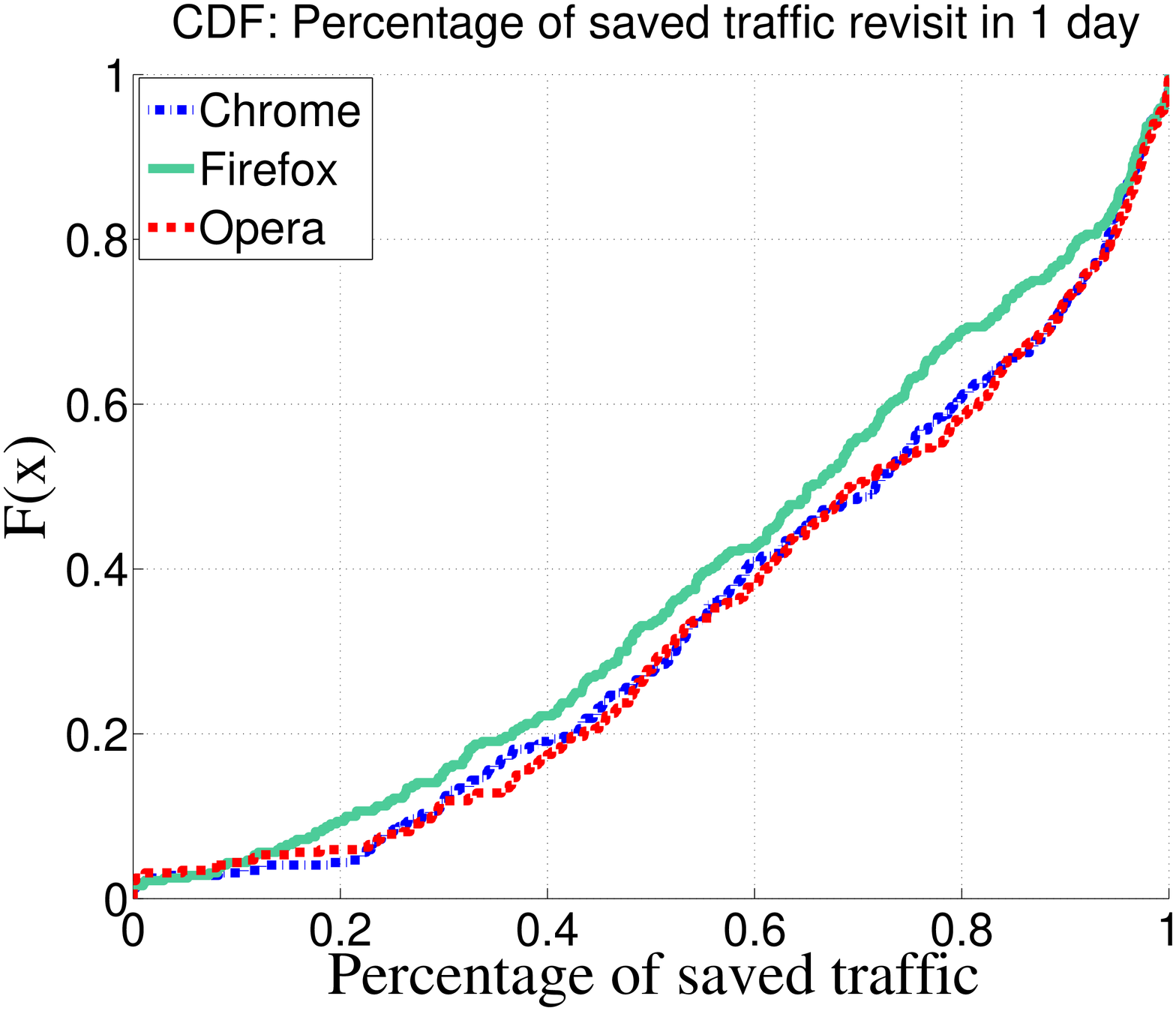}}
    \subfigure[]{
    \label{fig:realworld:e} 
    \includegraphics[width=0.32\textwidth]{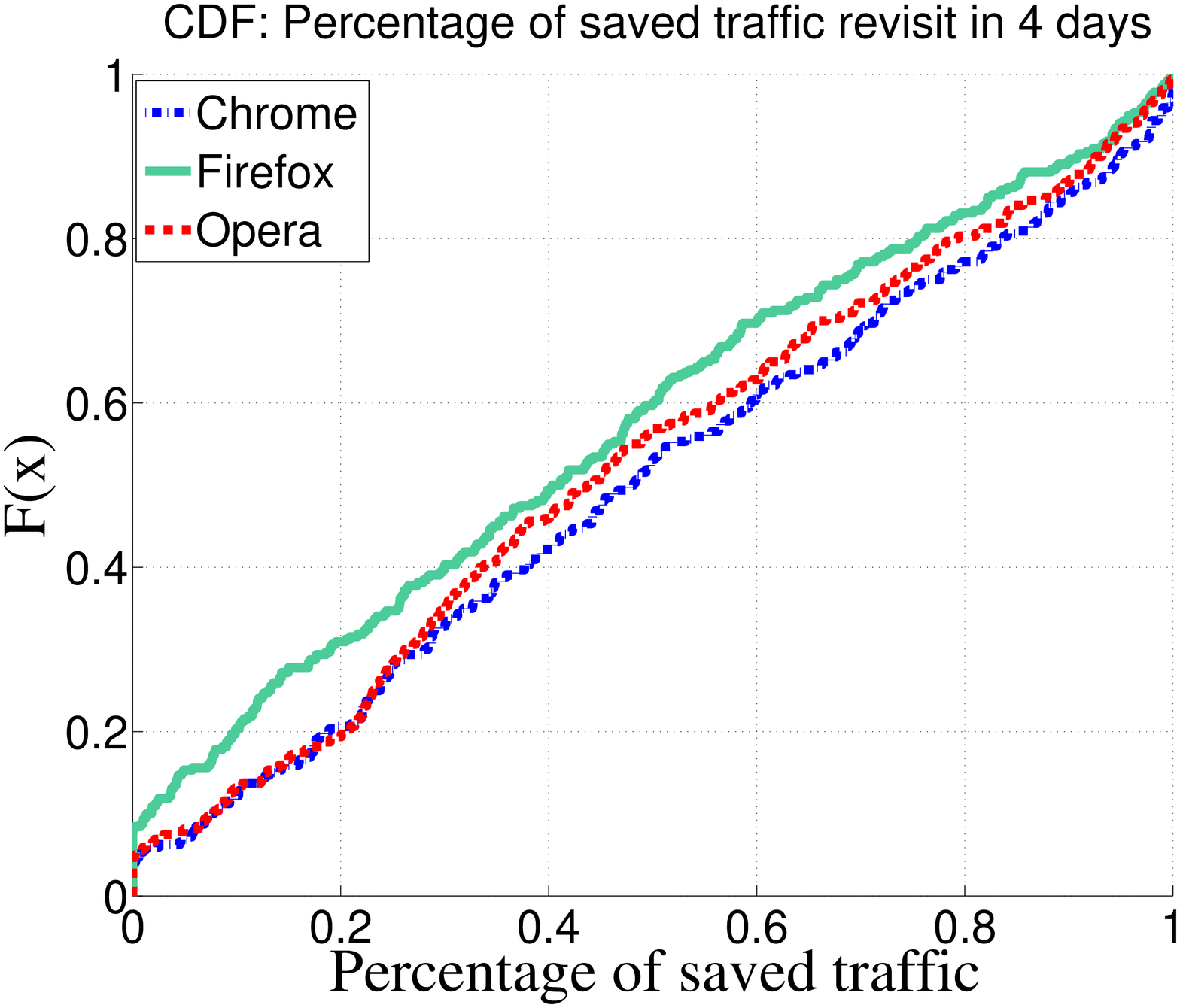}}
    \subfigure[]{
    \label{fig:realworld:f} 
    \includegraphics[width=0.32\textwidth]{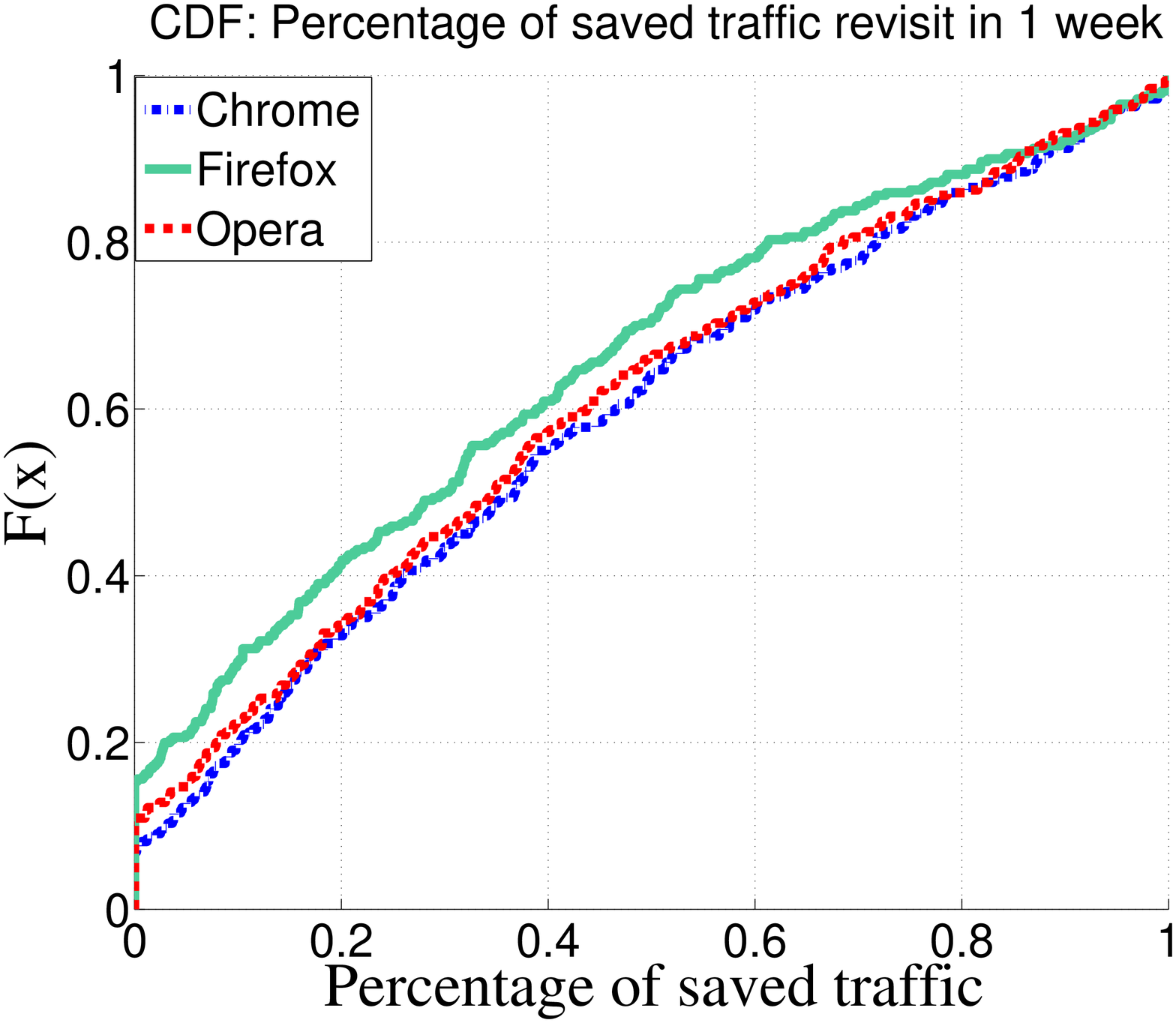}}
  \caption{CDF of the percentage of saved traffic. The x-axis is the median percentage of saved traffic of a website. The y-axis is the CDF. The revisiting interval from (a) to (g) is 30 minutes, 6 hours, 12 hours, 1 day, 4 days, 1 week. The red line is Firefox, blue is Chrome and brown is Opera.}
  \label{figure:realword} 
\end{figure*}

\subsection{Analysis of Uncached Resources}
We get a breif idea about the total cache hit rate through the data analysis as metioned above. The Firefox shows poorer cache performance. To take a more in-depth analysis, we classified the resources by their MIME type. We show the median percentage of HTML, image and JavaScript resources in the uncached resources.

We find that Opera has higher proportion of HTML resources that cannot be cached. Chrome has higher proportion of image resources that cannot be cached. Firefox has higher proportion of JavaScript resources that cannot be cached.

\begin{figure}[!t]
  \centering
\includegraphics[width=0.5\textwidth]{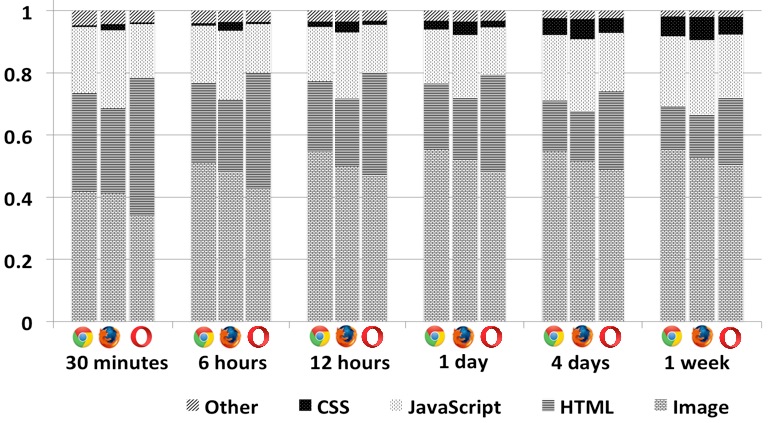}
  \caption{The percentage of different MIME type resources within the uncached resources. The x-axis denotes the revisiting interval. Within each interval, we show the compositon of uncached resources of Chrome, Firefox and Opera.}
  \label{figure:mime}
\end{figure}

Compared with HTML and image resources, JavaScript obviously are much more dynamic type. Large percentage of JavaScript explains why Firefox need to consume much more traffic data than others.

\begin{figure*}[!t]
  \centering
  \subfigure[]{
    \label{fig:realworld:a} 
    \includegraphics[width=0.32\textwidth]{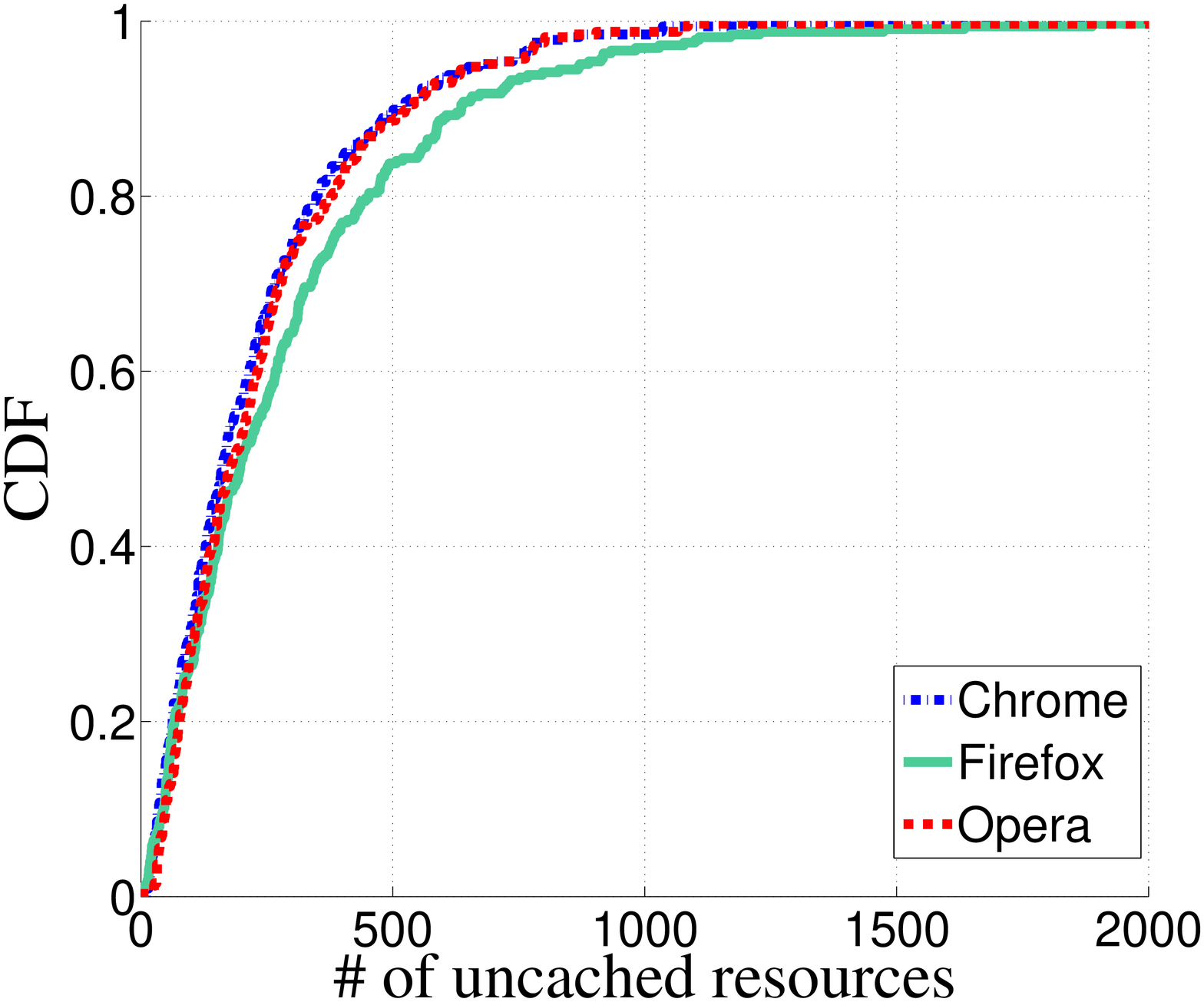}}
  \subfigure[]{
    \label{fig:realworld:b} 
    \includegraphics[width=0.32\textwidth]{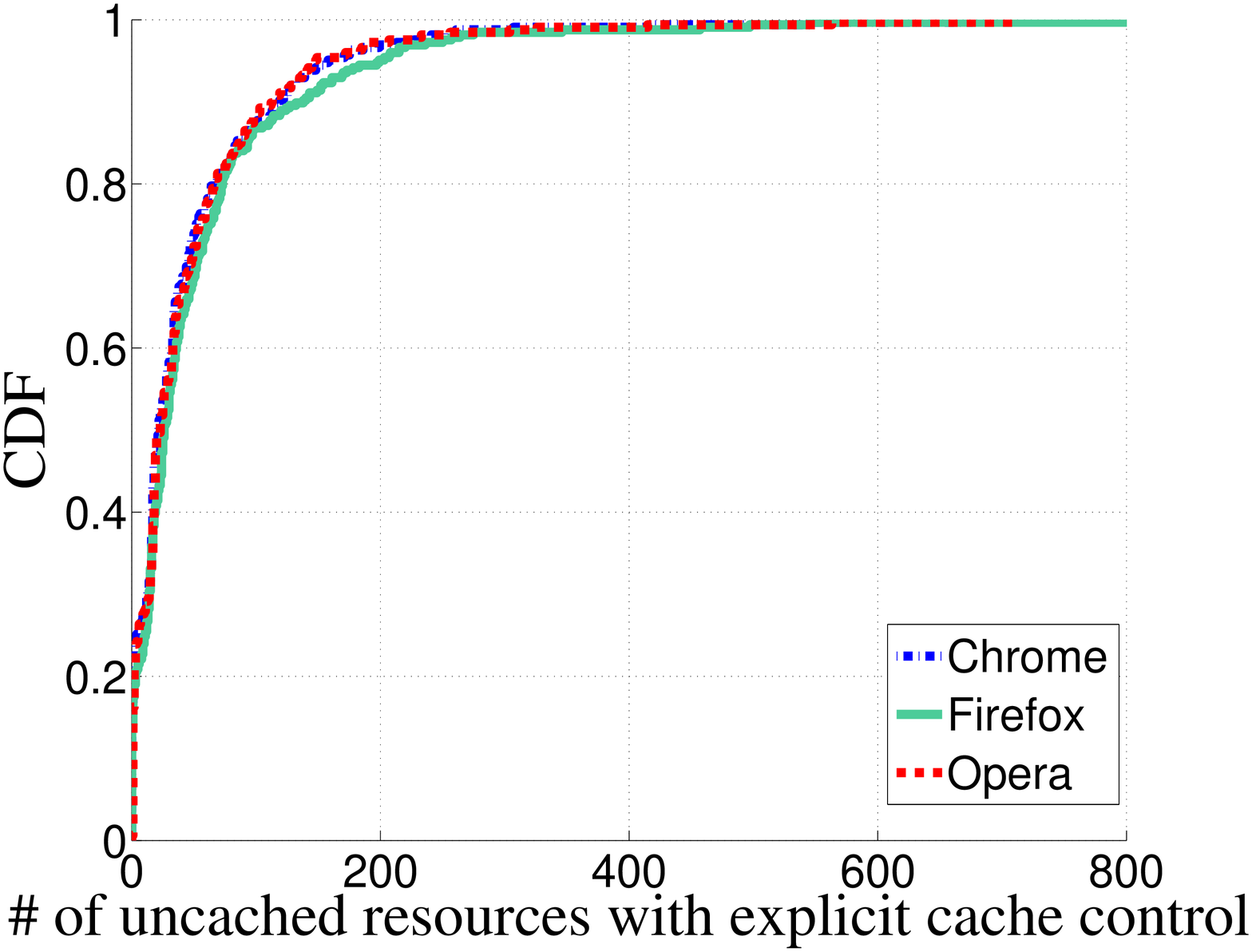}}
  \subfigure[]{
    \label{fig:realworld:c} 
    \includegraphics[width=0.32\textwidth]{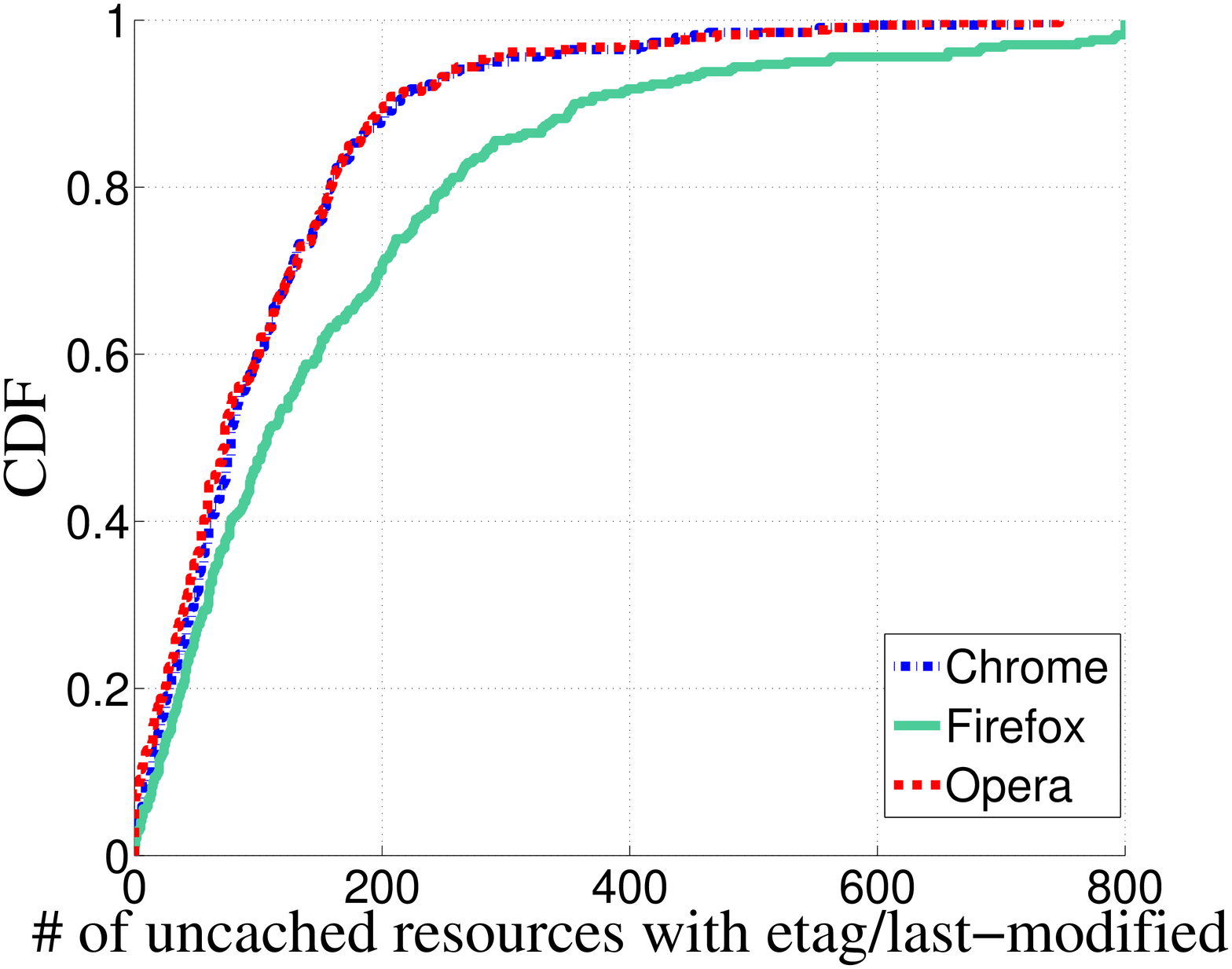}}
  \caption{The CDF of the numbers of uncached resources. (a) is the total numbers. (b) is the numbers of resources with explicit cache control.  (c) is the numbers of resources with etag or last-modified.}
  \label{figure:cachecontrol}
\end{figure*}

We analyze the cache performance from the perspective of explicit cache control in figure~\ref{figure:cachecontrol}. (a) is the CDF of total numbers of resources in uncached resources. (b) is the numbers of resources with explicit cache control in uncached resources. (c) is the numbers of resources with etag or last-modified in uncached resources. We can see that, although firefox have more numbers of resources cannot be cached, it actually shows advantage in the numbers of etag/last-modified resources. It seems to be a little bit confusing, because with more resources using etag/last-modified policy, Firefox should save data consumption, rather than waste data consumption. In our deeper research, we find the root cause.

Although Firefox has more resources using etag or last-modified policy, it actually always skips the validation and directly fetches the resource from the server. Here is an instance to show the problem of Firefox. When a client revisits the \emph{wikipedia.org}. It fetches a resource in every visit with the url of \url{https://bits.wikimedia.org/meta.wikimedia.org/load.php?debug=false&lang=en&modules=ext.gadget.wm-portal&only=styles&skin=vector&*}.
The resource is a css file and remains unchanged for a considerable time. The cache control policy of the resource is that \emph{public, max-age=300, s-maxage=300}, which means that the browser should keep the resource for 300 seconds. What's more, as the resource uses etag and last-modified, the revisits beyond 300 seconds should confirm whether the resource changes before fetching the whole resource.

Our experiment shows that, for Chrome and Opera, in an interval of 30 minutes, both browsers fetch the resource with the status code \emph{304}, which means both browsers fetch the resource from the local cache after validating the unchangeness of the content. While for firefox, it becomes complex and confusing. In some sitiations, Firefox fetches the resource with the status code \emph{304}. In other situations, Firefox fetches the resource with the status code \emph{200}, which means Firefox skips the validation and directly fetches the resource from the server. And this finally cause the poor cache performance of Firefox.
%
%

\begin{table*}[ht]
\centering
\footnotesize
\caption{Findings and implications}\label{table:implication}
\begin{tabular}{|p{8cm}|p{8cm}|}
\hline
\textbf{Findings} & \textbf{Implications}\\
\hline
The scale of the webpage fetched by different browsers are significantly different. About 90\% webpages have different scales among Chrome, Firefox and Opera.
&
Different browsers may convey quite different QoE for end-users. End-users should carefully choose browsers for different webpages.\\
\hline
Large percentage($>$60\%) of webpages have different proportion of HTML, image and JavaScript resources. As for CSS abd JSON, the percentage($<$30\%) of webpages is rather small.
&
Layouts and data of a webpage are more stable among different browsers while the structure and media objects may be more different.\\
\hline
Most webpages have similar metrics among Chrome and Opera than Firefox.
&
The kernel of Chrome and Opera is similar as they both use Blink while the kernel of Firefox is Gecko. However, although Chrome and Opera share the similar kernel, other internal logics, such as Javascript engine, may be different. \\
\hline
The common resources of all three browsers have a high proportion of image resources while have a low proportion of HTML resources. The distinct resources fetched by only one browser have a high proportion of HTML while have a low proportion of image.
&
Within all kinds of resources, HTML tends to be different among browsers while image tends to be the same. Behind the similar look-and-feel of webpage, the definition of layout can be various among browsers.\\
\hline
Firefox's cache performance is significantly poorer than Chrome and Opera. With the increase of revisiting interval, the gap between browsers become smaller.
&
For webpages that are revisited within a larger interval, there is no big difference on data consumption among different browsers. But for those frequently revisited webpages, it may be better to use Chrome or Opera.\\
\hline
Firefox requests more resources using \texttt{E-tag} or \texttt{last-modified} policy, but it actually always skips the validation and directly fetches the resource from the server.
&
Browser vendors should pay attention to the implementation of \texttt{E-tag} and \texttt{last-modified}, which can lead to undesired data consumption.\\
\hline
Opera has higher proportion of HTML resources that cannot be cached. Chrome has higher proportion of image resources that cannot be cached. Firefox has higher proportion of Javascript resources that cannot be cached.
&
Compared to HTML and image resources, Javascript can generate a lot of dynamic contents. Large percentage of Javascript involved in a webpage should be the reason why Firefox consume much more traffic data than others.\\
\hline
Webpages with large proportion of Javascript and lower proportion of images have shorter loading time on Chrome. Webpages with large proportion of images perform better on Firefox. Webpages with more third-party resources perform better on Opera.
&
The developers should make best use of the benefits and bypass the disadvantages of a browser. What's more, users can use different browsers to visit different kinds of webpages to reduce the data traffic.\\
\hline
Webpages with large percentage of HTTPS tend to have less loading time on Opera, while have more loading time on Firefox.
&
Opera developers should pay attention to their implementation of HTTPS. For secure connections, it's better to use Firefox.\\
\hline
Some kinds of MIME type resources, especially CSS and JSON, do not have much influence on the loading time among browsers.
&
The different logics of layouts on different browsers does not influence the loading time so that developer should consider more to ensure consistent look-and-feel.\\
\hline\end{tabular}
\end{table*}

\section{Implication}
\label{implication}
In this paper, we take a detailed analysis on the differences between three browsers: Chrome, Firefox and Opera. There are some interesting and meaningful results. In this section, we list the results and their implications in Table~\ref{table:implication}.
\section{Related Work}
\label{relatedwork}
\textbf{The cross-browser compatibility problem} has gained a lot of research efforts since the birth of the Web browser. It is widely recognized as an important issue among Web developers but hardly ever addressed directly during the software development process~\cite{Rode:TR2005,Liu:TSC09}\cite{ICSE16Lu}\cite{TSC16Huang}.
S. R. Choudhary and M. R. Prasad and A. Orso~\cite{Choudhary:ICSE2013} find that, due to the increasing popularity of web applications, and the number of browsers and platforms on which such applications can be executed, cross-browser incompatibilities (XBIs) are becoming a serious concern for organizations that develop web-based software
Choudhary et al.~\cite{Choudhary:ICSM2010} proposed a technique for automatically detecting cross-browser issues and assisting their diagnosis. Given a page to be analyzed, the comparison is performed by combining a structural analysis
of the information in the page`s DOM and a visual analysis of the page`s appearance, obtained through screen captures.
Mesbah and Prasad~\cite{Mesbah:ICSE2011} defined the cross-browser compatibility problem and proposed a systematic, fully-automated approach for cross-browser compatibility testing that can expose a substantial fraction of the cross-browser issues in modern dynamic Web applications. A navigation model is used to compare two Web applications based on trace equivalence and screen equivalence.
To try to solve the cross-browser compatibility problem S. R. Choudhary~\cite{Choudhary:ICSE2011} builds WEBDIFF, the first technique to apply concepts from computer vision and graph theory to identify cross-browser issues in web applications.

\textbf{Research about the website characterization} is an important research question. A lot of work have done to characterize and describe the factors of web related problem~\cite{Li:IMC15}~\cite{MorleyMao:IMC2010}~\cite{qian:2010}~\cite{Qian:Mobsys2014}~\cite{Huang:IMC2012}.
Fred Douglis, Anja Feldmann, Balachander Krishnamurthy and Jeffrey C. Mogul~\cite{Douglis:USITS1997} take research about characteristics of Web resources, including access rate, age at time of reference, content type,
resource size, and Internet top-level domain.
Butkiewicz, Michael and Madhyastha, Harsha V. and Sekar, Vyas~\cite{Butkiewicz:SIGCOMM2011} identify a set of metrics to characterize the complexity of websites both at a content-level (e.g., number and size of images) and service-level (e.g., number of servers/origins).They find that the distributions of these metrics are largely independent of a website's popularity rank. However, some categories (e.g., News) are more complex than others.
Those researches help to characterize the content of web pages by several interesting metrics.

\textbf{The analysis of the mobile web performance} is an important research question in the past years.
Huang \textit{et al.}~\cite{Huang:2010}measured mobile Web browsing performance.
Wang \textit{et al.}~\cite{ZhongLin:2011} examined the issues specific to mobile Web browsers. Mobile HTTP Archive~\cite{MobileHttpArchive} records mobile Web performance information of about 5000 mobile websites. But its recording period is 15 days, and is too coarse-grained to analyze the cache performance.
Qian et al.~\cite{MorleyMao:MobiSys2012} perform the first network-wide study of the redundant transfers caused by inefficient Web caching on handsets.
Niranjan Balasubramanian, Aruna Balasubramanian et al~\cite{Niranjan:SIGCOMM2009}. present a measurement study of the energy consumption characteristics of three widespread mobile networking technologies: 3G, GSM, and WiFi.
Qian et al~\cite{Qian:Mobsys2014} take a first comprehensive examination of the resource usage of mobile Web browsing by focusing on two important types of resources: bandwidth and energy.
Our previous work studied the performance of JavaScripts in mobile Web browsing~\cite{TOIT17Liu}, and the comparison between native apps and Web apps~\cite{ICWS15Liu}.

\textbf{A lot of efforts have been invested to study the performance of mobile Web cache}.
Wang \textit{et al.}~\cite{ZhongLin:WWW2012} examined three client-only solutions to accelerate mobile browser speed: caching, prefetching, and speculative loading, by using Web usage data collected from 24 iPhone users over one year. They found that caching has very limited effectiveness: 60\% of the requested resources are either expired or not in the cache.
Zhang \textit{et al.}~\cite{ZhangYiFan:UbiComp2013} performed a comprehensive measurement study on Web caching functionality of 1300 top ranked Android apps, not just Web browsers. Results revealed that imperfect web caching is a common and serious problem for Android apps generating Web traffic. They also implemented a system-wide service called CacheKeeper to effectively reduce overhead caused by poor Web caching of mobile apps.
Qian \textit{et al.}~\cite{MorleyMao:MobiSys2012} conducted a measurement study on Web caching in smartphones. By examining a one-day smartphone Web traffic dataset collected from a cellular carrier and a five-month Web access trace collected from 20 smartphone users, the study revealed that about 20\% of the total Web traffic examined is redundant due to imperfect cache implementations.
Our previous work \textit{et al.}~\cite{WWW15Ma}\cite{TMC16Liu} studied the mobile Web cache performance. In our following work~\cite{TMC16Liu}\cite{TMC16Liu}, we optimize the cache performance.

\textbf{Measurement of resource loading} is also an important research question.
Wang et al.~\cite{Wang:HotMobile2011} advocated that resource loading contributes most to the browser delay.
Wang et al.~\cite{Wang:NSDI2013} designed a lightweight in-browser profiler, called WProf, and studied the dependencies of activities when browsers load a webpage.
Nejati et al.~\cite{Nejati:WWW2016} extended WProf to WProf-M and studied the differences of page loading process between mobile and non-mobile browsers.
Li et al.~\cite{Li:NSDI2010} designed WebProphet to capture dependencies among Web resources and to automate the prediction of user-perceived Web performance.

However, none of the work cast light on the differences from the perspective of browsers. Our work takes discussion about the mobile performance, i.e, cache and loading from the view of browsers.

\section{Conclusions}
\label{conclusion}
In this paper, we conduct a comprehensive study on the quality of experiences among different mobile browsers. We design and establish an automated data-collection platform to collect resources from the same webpages when accessed from different browsers for a considerably long time. We choose Chrome, Firefox and Opera as our target browsers. We use several metrics to compare the differences of resources acquired from different browsers. Based on the resource characteristics, we further analyze the loading time and cache performance of different browsers, revealing the root causes. Our findings can benefit browser vendors, web developers as well as end-users.

\bibliographystyle{abbrv}
\bibliography{imc16,mobisaas}

\begin{thebibliography}{10}

\bibitem{Alexa}
Alexa.
\newblock http://www.alexa.com/.

\bibitem{Charles}
Charles proxy.
\newblock http://www.charlesproxy.com/.

\bibitem{UCWeb}
Iconic cloud acceleration and data compression.
\newblock http://www.ucweb.com/a/ppress/2015/0713/4314.html.

\bibitem{browCNZZ}
Market share of browsers of smartphones in china.
\newblock http://brow.data.cnzz.com/.

\bibitem{MobileHttpArchive}
Mobile {HTTP} archive.
\newblock http://mobile.httparchive.org/.

\bibitem{3rdwebview}
Third party webviews on android.
\newblock
  http://infil00p.org/android/cordova/2014/03/14/third-part-webviews-on-android/.

\bibitem{Niranjan:SIGCOMM2009}
N.~Balasubramanian, A.~Balasubramanian, and A.~Venkataramani.
\newblock Energy consumption in mobile phones: a measurement study and
  implications for network applications.
\newblock In {\em Proceedings of the 9th ACM SIGCOMM conference on Internet
  measurement conference}, pages 280--293, 2009.

\bibitem{Butkiewicz:SIGCOMM2011}
M.~Butkiewicz, H.~V. Madhyastha, and V.~Sekar.
\newblock Understanding website complexity: Measurements, metrics, and
  implications.
\newblock In {\em Proceedings of the 2011 ACM SIGCOMM Conference on Internet
  Measurement Conference}, pages 313--328, 2011.

\bibitem{Choudhary:ICSE2011}
S.~R. Choudhary.
\newblock Detecting cross-browser issues in web applications.
\newblock In {\em Proceedings of the 33rd International Conference on Software
  Engineering, {ICSE} 2011}, pages 1146--1148.

\bibitem{Choudhary:ICSE2013}
S.~R. Choudhary, M.~R. Prasad, and A.~Orso.
\newblock X-pert: Accurate identification of cross-browser issues in web
  applications.
\newblock In {\em Proceedings of the International Conference on Software
  Engineering, {ICSE} 13}, pages 702--711, 2013.

\bibitem{Choudhary:ICSM2010}
S.~R. Choudhary, H.~Versee, and A.~Orso.
\newblock {WEBDIFF}: Automated identification of cross-browser issues in web
  applications.
\newblock In {\em Proceedings of the 2010 IEEE International Conference on
  Software Maintenance}, pages 1--10, 2010.

\bibitem{Douglis:USITS1997}
F.~Douglis, A.~Feldmann, B.~Krishnamurthy, and J.~C. Mogul.
\newblock Rate of change and other metrics: a live study of the world wide web.
\newblock In {\em Proceedings of the 1997 USITS conference {USITS'}97}, 1997.

\bibitem{Qian:Mobsys2014}
Q.~Feng, S.~Sen, and O.~Spatscheck.
\newblock Characterizing resource usage for mobile web browsing.
\newblock In {\em Proceedings of the 12th annual international conference on
  Mobile systems, applications, and services}, pages 218--231, 2014.

\bibitem{TSC16Huang}
G.~Huang, X.~Liu, X.~Lu, Y.~Ma, Y.~Zhang, and Y.~Xiong.
\newblock Programming situational mobile web applications with cloud-mobile
  convergence: An internetware-oriented approach.
\newblock {\em IEEE Transactions on Services Computing}, 2016.

\bibitem{Huang:IMC2012}
J.~Huang, F.~Qian, Z.~M. Mao, S.~Sen, and O.~Spatscheck.
\newblock Screen-off traffic characterization and optimization in 3g/4g
  networks.
\newblock In {\em Proceedings of the 2012 ACM conference on Internet
  measurement conference, {IMC} 2012}, pages 357--364. ACM, 2012.

\bibitem{Huang:2010}
J.~Huang, Q.~Xu, B.~Tiwana, Z.~M. Mao, M.~Zhang, and P.~Bahl.
\newblock Anatomizing application performance differences on smartphones.
\newblock In {\em Proceedings of the 2010 international conference on Mobile
  systems, applications, and services, {MobiSys} 2010}, pages 165--178, 2010.

\bibitem{Li:IMC15}
H.~Li, X.~Lu, X.~Liu, T.~Xie, K.~Bian, F.~X. Lin, Q.~Mei, and F.~Feng.
\newblock Characterizing smartphone usage patterns from millions of {Android}
  users.
\newblock In {\em Proceedings of the ACM SIGCOMM Conference on Internet
  Measurement, {IMC} 15}, pages 459--472, 2015.

\bibitem{Li:NSDI2010}
Z.~Li, M.~Zhang, Z.~Zhu, Y.~Chen, A.~Greenberg, and Y.-M. Wang.
\newblock Webprophet: Automating performance prediction for web services.
\newblock In {\em Proceedings of the 7th {USENIX} conference on Networked
  systems design and implementation, {NSDI} 2010}, pages 143--158, 2010.

\bibitem{TOIS17Liu}
X.~Liu, W.~Ai, H.~Li, J.~Tang, G.~Huang, F.~Feng, and Q.~Mei.
\newblock Deriving user preferences of mobile apps from their management
  activities.
\newblock {\em ACM Transactions on Information Systems (TOIS)}, 35(4):39, 2017.

\bibitem{Liu:TSC09}
X.~Liu, G.~Huang, and H.~Mei.
\newblock Discovering homogeneous web service community in the user-centric web
  environment.
\newblock {\em {IEEE} Trans. Services Computing}, 2(2):167--181, 2009.

\bibitem{TSE17Liu}
X.~Liu, H.~Li, X.~Lu, T.~Xie, Q.~Mei, H.~Mei, and F.~Feng.
\newblock Understanding diverse smartphone usage patterns from large-scale
  appstore-service profiles.
\newblock {\em IEEE Transactions on Software Engineering}, page Accepted to
  appear, 2017.

\bibitem{TMC16Liu}
X.~Liu, Y.~Ma, Y.~Liu, X.~Wang, T.~Xie, and G.~Huang.
\newblock Swarovsky: Optimizing resource loading for mobile web browsing.
\newblock {\em IEEE Transactions on Mobile Computing}, 2016.

\bibitem{TOIT17Liu}
X.~Liu, Y.~Ma, M.~Yu, Y.~Liu, G.~Huang, and H.~Mei.
\newblock i-jacob: An internetware-oriented approach to optimizing
  computation-intensive mobile web browsing.
\newblock {\em {ACM} Trans. Internet Techn.}, page Accepted to appear, 2017.

\bibitem{ICWS15Liu}
Y.~Liu, X.~Liu, Y.~Ma, Y.~Liu, Z.~Zheng, G.~Huang, and M.~B. Blake.
\newblock Characterizing restful web services usage on smartphones: {A} tale of
  native apps and web apps.
\newblock In {\em 2015 {IEEE} International Conference on Web Services, {ICWS}
  2015, New York, NY, USA, June 27 - July 2, 2015}, pages 337--344, 2015.

\bibitem{ICSE16Lu}
X.~Lu, X.~Liu, H.~Li, T.~Xie, Q.~Mei, D.~Hao, G.~Huang, and F.~Feng.
\newblock {PRADA:} prioritizing android devices for apps by mining large-scale
  usage data.
\newblock In {\em Proceedings of the 38th International Conference on Software
  Engineering, {ICSE} 2016, Austin, TX, USA, May 14-22, 2016}, pages 3--13,
  2016.

\bibitem{WWW15Ma}
Y.~Ma, X.~Liu, S.~Zhang, R.~Xiang, Y.~Liu, and T.~Xie.
\newblock Measurement and analysis of mobile web cache performance.
\newblock In {\em Proceedings of the 24th International Conference on World
  Wide Web, {WWW} 2015, Florence, Italy, May 18-22, 2015}, pages 691--701,
  2015.

\bibitem{Mesbah:ICSE2011}
A.~Mesbah and M.~R. Prasad.
\newblock Automated cross-browser compatibility testing.
\newblock In {\em Proceedings of the 33rd International Conference on Software
  Engineering, {ICSE} 2011}, pages 561--570, 2011.

\bibitem{Nejati:WWW2016}
J.~Nejati and A.~Balasubramanian.
\newblock An in-depth study of mobile browser performance.
\newblock In {\em Proceedings of the 25th International Conference on World
  Wide Web, {WWW} 2016}, 2016.

\bibitem{MorleyMao:MobiSys2012}
F.~Qian, K.~S. Quah, J.~Huang, J.~Erman, A.~Gerber, Z.~Mao, S.~Sen, and
  O.~Spatscheck.
\newblock Web caching on smartphones: ideal vs. reality.
\newblock In {\em Proceedings of the 2012 International Conference on Mobile
  Systems, {MobiSys}2012}, pages 127--140, 2012.

\bibitem{MorleyMao:IMC2010}
F.~Qian, Z.~Wang, A.~Gerber, Z.~M. Mao, S.~Sen, and O.~Spatscheck.
\newblock Characterizing radio resource allocation for {3G} networks.
\newblock In {\em Proceedings of the 2010 ACM conference on Internet
  measurement conference, {IMC} 2010}, pages 137--150, 2010.

\bibitem{qian:2010}
F.~Qian, Z.~Wang, A.~Gerber, Z.~M. Mao, S.~Sen, and O.~Spatscheck.
\newblock Characterizing radio resource allocation for 3g networks.
\newblock In {\em Proceedings of the 10th ACM SIGCOMM conference on Internet
  measurement}, pages 137--150, 2010.

\bibitem{Rode:TR2005}
J.~Rode, M.~B. Rosson, and M.~A. P{\'e}rez-Qui{\~n}ones.
\newblock The challenges of web engineering and requirements for better tool
  support.
\newblock 2005.

\bibitem{Wang:NSDI2013}
X.~S. Wang, A.~Balasubramanian, A.~Krishnamurthy, and D.~Wetherall.
\newblock Demystifying page load performance with {WProf}.
\newblock In {\em Proceedings of {USENIX} Conference on Networked Systems
  Design and Implementation, {NSDI} 2013}, pages 473--485, 2013.

\bibitem{ZhongLin:2011}
Z.~Wang, F.~X. Lin, L.~Zhong, and M.~Chishtie.
\newblock Why are web browsers slow on smartphones?
\newblock In {\em Proceedings of the 2011 International Conference on World
  Wide Web, {WWW} 2011}, pages 91--96, 2011.

\bibitem{Wang:HotMobile2011}
Z.~Wang, F.~X. Lin, L.~Zhong, and M.~Chishtie.
\newblock Why are web browsers slow on smartphones?
\newblock In {\em Proceedings of the 12th Workshop on Mobile Computing Systems
  and Applications, {HotMobile} 2011}, pages 91--96, 2011.

\bibitem{ZhongLin:WWW2012}
Z.~Wang, F.~X. Lin, L.~Zhong, and M.~Chishtie.
\newblock How far can client-only solutions go for mobile browser speed?
\newblock In {\em Proceedings of the 2012 International Conference on World
  Wide Web, {WWW} 2012}, pages 31--40, 2012.

\bibitem{ZhangYiFan:UbiComp2013}
Y.~Zhang, C.~Tan, and L.~Qun.
\newblock {CacheKeeper}: a system-wide web caching service for smartphones.
\newblock In {\em Proceedings of the 2013 ACM International Joint Conference on
  Pervasive and Ubiquitous Computing, {UbiComp} 2013}, pages 265--274, 2013.

\end{thebibliography}
\end{document}